# A multi-level collaborative filtering method that improves recommendations

Nikolaos Polatidis *, Christos K. Georgiadis

Department of Applied Informatics, University of Macedonia, 54006, Thessaloniki, Greece

*Corresponding Author. Tel +302310891810

Email addresses: npolatidis@uom.edu.gr (N. Polatidis), geor@uom.edu.gr (C.K. Georgiadis)

**Abstract**

Collaborative filtering is one of the most used approaches for providing recommendations in various online environments. Even though collaborative recommendation methods have been widely utilized due to their simplicity and ease of use, accuracy is still an issue. In this paper we propose a multi-level recommendation method with its main purpose being to assist users in decision making by providing recommendations of better quality. The proposed method can be applied in different online domains that use collaborative recommender systems, thus improving the overall user experience. The efficiency of the proposed method is shown by providing an extensive experimental evaluation using five real datasets and with comparisons to alternatives.

**Keywords:** Collaborative Filtering, Similarity, Multi-level, Hybrid, Recommender System

## 1. Introduction

Nowadays, more and more people find that the constant growth of the web in combination with the development of technologies such as smartphones and tablets results in spending more time accessing information online. However, these developments have brought a massive amount of information, resulting in an information overload problem (Moradi & Ahmadian, 2015; Bobadilla et al., 2013). With too much information all over the web, users find it very challenging to find the data they need. For that reason most users find it frustrating when looking online for what movie to watch, who to add as a friend in a social network and many other related search problems. The solution to this problem is recommender systems, which apply techniques developed to analyze user data and make recommendations that the user will probably like. It is a method that both the service provider and the user are benefited from. The main reasons for the mutual benefits include the fast processing of data for the service provider, the higher percentage of sales, the saving of time for the user and the discovery of products or services that otherwise would be difficult to find (Polatidis & Georgiadis, 2013).

Collaborative filtering is the most known and widely used technique for providing fast and accurate enough recommendations to users (Shi et al., 2014; Ekstrand et al., 2011; Konstan & Riedl 2012; Su & Khoshgoftaar 2009). This method relies on a database of ratings submitted by each user for products or services, then the ratings are compared to each other with the use of suitable similarity method in order to provide recommendations to the user who makes the request. The main two functions of such systems is to identify a pre-specified number of neighbors according to similar ratings and then provide the recommendations. Collaborative filtering has been widely adopted by many real-world systems, such as Netflix and Amazon (Wang et al., 2015), and this is due to its simplicity and efficiency.

In addition to collaborative filtering, other recommendations methods include content-based, which is based on item metadata. In this method the user supplies a set of information and preferences and the algorithm makes recommendations according to the settings provided (Bobadilla et al., 2013) and Burke (2007). Moreover, knowledge-based recommender is another recommendation approach that uses inferences about user preferences and specific knowledge about the domain and also how the items or services to be recommended meet the preferences set by the users (Jannach et al., 2010). A widely used recommendation approach is the combination of one or more recommendation methods, and is called hybrid (Burke, 2002). It doesn't necessarily mean that the two methods must be different, but they could be two different collaborative filtering methods as well (Jannach et al. 2010; Burke, 2007).

This paper proposes the use of a new recommendation collaborative filtering method that aims to improve the accuracy of the recommendations. In most collaborative recommendation systems a similarity function, such as Pearson Correlation Coefficient (PCC) or Cosine (Shi et al., 2014), is utilized by the system to provide recommendations by taking in consideration the absolute ratings between users. Our motivation is to divide user similarity, as offered by PCC, into different levels and add constraints to each level. We show that by modifying the user similarity, which is a value from -1 to 1, according to the constraints that each user belongs, the accuracy of the recommendations is improved. Furthermore, we argue that the quality of the recommendations is improved as well when the constraints are at place. The proposed method attempts to provide recommendations of better accuracy and quality when compared to other alternatives. However for this to be done correctly, enough ratings should have already been submitted by users of the system.

The contributions of the paper are:

1. We propose a recommendation method that improves the accuracy of collaborative filtering and is based on multiple levels and constraints.

2. We perform extensive experiments using five real datasets in order to evaluate our proposed method and compare it against alternatives in order show that our proposed method is both practical and effective.

The rest of the paper is organized as follows: section 2 is the related work part, section 3 describes the proposed method, section 4 explains the experimental evaluation and section 5 contains the conclusions and future work part.

## 2. Related work

While collaborative filtering methods have been widely used by many real world systems, including Netflix and Amazon, there are not sufficient details available regarding the provided recommendations. Collaborative filtering techniques use a database of ratings among users and items, such as the one shown in table 1, must be present (Shi et al., 2014).

| | Item/Service 1 | Item/Service 2 | Item/Service 3 | Item/Service 4 | Item/Service 5 |
|---|---|---|---|---|---|
| User 1 | - | 2 | 3 | 5 | - |
| User 2 | 1 | 2 | 4 | 5 | 5 |
| User 3 | 2 | 5 | 1 | 1 | 2 |
| User 4 | 3 | - | - | - | 3 |
| User 5 | - | 3 | 4 | - | - |

**Table 1.** A database of ratings

When recommendations need to be generated for a user, then the ratings are loaded into memory and a similarity function is used. The main part is how to estimate the similarity value between two users. This is called neighborhood identification and the job of the similarity function is to firstly identify a pre-specified of $k$ nearest neighbors according to their similarity value. In present recommendations systems the value of $k$ can vary from a few, possibly 2 to 5, to as many as possible with the number ranging from 10 to 20 to 30 and so on up to hundreds of neighbors. A high number of neighbors doesn't necessarily mean that the accuracy of the recommendations will be high though.

Now, as mentioned for the identification of the nearest neighbors a similarity function such as PCC is necessary to be used. PCC is defined in equation 1. In PCC the sum of ratings between two users is compared. Sim (a, b) is the similarity between users a and b, also $r_{a,p}$ is the rating of user a for product p, $r_{b,p}$ is the rating of user b for product p and $\bar{r}a$ and $\bar{r}b$ represent the user's average ratings. P is the set of all products. Moreover, the similarity value ranges from -1 to 1 and higher is better.

$$Sim_{a,b}^{PCC} = \frac{\sum p \in P(ra,p - \bar{r}a)(rb,p - \bar{r}b)}{\sqrt{\sum p \in P(ra,p - \bar{r}a)^2}\sqrt{\sum p \in P(rb,p - \bar{r}b)^2}} \quad (1)$$

After the similarity values are computed according to the equation used and the formation of the $k$ nearest neighborhood, then the rating procedure takes place, where ratings are being predicted for items. The items with the highest rating predicted value are being recommended to the user who made the request. On the other hand, PCC has inspired other similarity methods such as the weighted PCC (WPCC) (Herlocker et al., 1999) defined in equation 2. WPCC is based on PCC and provides recommendations based additionally on the number $T$ of co-rated items between users. In their work the value was set to 50, which means that if the number of the co-rated items was 50 or more the recommendations from these users are preferred. Also in the case that less ratings are available then the algorithm switches to ordinary PCC.

$$Sim_{a,b}^{WPCC} = \left\{ \begin{matrix} \frac{|Ia \cap Ib|}{T} \cdot Sim_{a,b}^{PCC}, & if \left|Ia \bigcap Ib\right| < T \\ Sim_{a,b}^{PCC}, & otherwise \end{matrix} \right\} \quad (2)$$

A somewhat similar approach for identifying neighbors is proposed by Jamali & Ester (2009) and is defined in equation 3. In this method the similarity of small number of co-rated items is weaken.

$$Sim_{a,b}^{SPCC} = \frac{1}{1 + \exp(-|Ia \cap Ib|/2)} \cdot Sim_{a,b}^{PCC} \quad (3)$$

Another approach to recommendations based on collaborative filtering is Jaccard similarity (Koutrika et al., 2009). In this approach the similarity computation of PCC is not used, but only the number of co-rated items is taken into consideration. Jaccard similarity is defined in equation 4.

$$Sim_{a,b}^{Jaccard} = \frac{|Ia \cap Ib|}{|Ia \cup Ib|} \quad (4)$$

Other proposed similarities for collaborative filtering include the mean squared difference (MSD) (Cacheda et al., 2011). This method captures the difference that the users have in their ratings. Furthermore, another proposed similarity measure has been proposed by Lu et al. (2013) where the use of fuzzy set theory is used with the aim to assign different weight values to different rating differences. Another method, proposed by Wang et al. (2015), uses entropy to provide user similarity in collaborative filtering. In this work the majority of ratings used by PCC must be similar. Liu et al. (2014) proposed a collaborative filtering improvement that does not only consider the local user rating information, but also the global behavior of the user. Son (2014) proposed a fuzzy recommendation method that uses demographic data instead of user ratings. One more similarity measure found in the literature is proposed by Bobadilla et al. (2012a). This recommendation method tries to solve the cold start problems found in recommender systems. In particular it aims to solve the problem of new users stop using the system because of low accuracy found at the initial stages (when just a few ratings are available). For this reason, the authors propose a new similarity measure based on neural learning and uses optimization that provides better accuracy to users with few submitted ratings. An alternative similarity measure based on singularities is described in Bobadilla et al. (2012b). In this approach contextual information derived from users is used to calculate the singularity for each item. According to their results the similarity is improved when compared to simple collaborative filtering. Ahn (2008) offers, yet, another similarity measure that alleviates the new user cold start problem. This paper proposes a heuristic measure that improves the similarity under cold start scenarios, and in particular when only few ratings are available. Anand (2011) proposed a recommendation method that utilizes a number of sparsity measures based on both local and global similarities and it is still another method that improves the accuracy of recommendations that are based on collaborative filtering. Moreover, recommendation methods that aim to improve the accuracy of collaborative recommendations but the approach offered by Moradi and Ahmadian (2015) is different, since it uses data from both ratings submitted from users and from social-trust networks that are part of the recommendation system. Also, another approach that uses not only trust but distrust information as well, is the one proposed by Fang et al. (2015). Moreover, Polatidis and Georgiadis (2015) offer an improved collaborative filtering similarity that is based on data from social rating networks and also show how this method can be used in ubiquitous environments and also preserve the user privacy if necessary. Toledo et al. (2015) proposed a recommendation method that aims to detect and correct unreliable ratings that could bias the recommendations provided and in order to do it the method characterizes the users from their profiles. Finally, it should be noted that Cosine similarity is used also widely by collaborative recommender systems and its main difference with PCC is that it does not take into consideration user habits with extremely low or high ratings (Shi at al., 2014; Ekstrand et al., 2011).

## 3. Proposed Method

In this section we give a description of our proposed method. The first step is to introduce our multi-level function, that aims to improve the accuracy of the recommendations, and then to address the problem where our function is not able to provide recommendations for certain users.

### 3.1 The steps of the proposed method

PCC is the most widely used and accurate similarity measure used by collaborative filtering recommender systems (Wang et al., 2015). Our goal is the enhancement of the PCC similarity measure, which results in the improvement of the accuracy performance. The difference among the ratings of the items that users have co-rated is the value of similarity that exists between these users. Many similarity measures get better accuracy by manipulating how these calculations take place and the methods are based on absolute rating differences (Wang et al., 2015). According to this approach, it is crucial that all of the values of co-rated items between two users must be the same. See table 2 for an example where it is shown that 'user 2 new' has a different value for item 5. This means that there is no similarity between 'user 2 new' and 'user 1' (while 'user 2' is similar to 'user 1').

| | User 1 | User 2 | User 2 new |
|---|---|---|---|
| Item 1 | 5 | 5 | 5 |
| Item 2 | 4 | 4 | 4 |
| Item 3 | 3 | 3 | 3 |
| Item 4 | 2 | 2 | 2 |
| Item 5 | 1 | 1 | 5 |

**Table 2.** Absolute ratings

While other methods use the method of absolute ratings either by adding weights, or by manipulating different variables to achieve a better result, our approach is based on a multi-level division. Furthermore, we argue that every user has a rating expression that should not be punished that way. Also, we agree with PCC based similarity approaches that consider higher the similarity when the values of the co-rated items are as close as possible. Our proposed method heavily relies on PCC, the similarity value returned from it and the number of co-rated items. In order to achieve this we first introduce a similarity function which is defined in equation 5. In this equation, $T$ stands for the total number of co-rated items, $x$ is a positive real number such as $x \in R$, $y$ is a positive real number such as $y \in R$. In our proposed method we argue that the by dividing the algorithm in multiple levels the accuracy of the recommendations is improved. To show the effectiveness of our method in this we made experiments using four levels. In this context, at the first four steps the number of co-rated items is checked and if it is more than the pre-specified thresholds (*t1, t2, t3, t4)*, then it can proceed to the next step to check the similarity value derived for the two users from PCC. For users that a sufficient number of co-rated items exist and the PCC similarity value is greater than a pre-specified threshold ($y$) then a list of recommendations is returned. Otherwise, for users that there is not available a sufficient number of co-rated items, zero value is returned. Finally, *t1, t2, t3* and *t4* are natural numbers that represent the constraints put on the number of co-rated items for each level ($t1 \in N$, $t2 \in N$, $t3 \in N$, $t4 \in N \wedge t4>t3>t2>t1$)

$$Sim\begin{matrix}Proposed\\a,b\end{matrix} = \begin{cases} Sim\begin{matrix}PCC\\a,b\end{matrix} + x, & if \frac{|Ia \cap Ib|}{T} \geq t1 \; and \; Sim\begin{matrix}PCC\\a,b\end{matrix} \geq y \\ Sim\begin{matrix}PCC\\a,b\end{matrix} + x, & if \frac{|Ia \cap Ib|}{T} < t1 \; and \; \frac{|Ia \cap Ib|}{T} \geq t2 \; and \; Sim\begin{matrix}PCC\\a,b\end{matrix} \geq y \\ Sim\begin{matrix}PCC\\a,b\end{matrix} + x, & if \frac{|Ia \cap Ib|}{T} < t2 \; and \; \frac{|Ia \cap Ib|}{T} \geq t3 \; and \; Sim\begin{matrix}PCC\\a,b\end{matrix} \geq y \\ Sim\begin{matrix}PCC\\a,b\end{matrix} + x, & if \frac{|Ia \cap Ib|}{T} < t3 \; and \; \frac{|Ia \cap Ib|}{T} \geq t4 \; and \; Sim\begin{matrix}PCC\\a,b\end{matrix} \geq y \\ 0, & otherwise \end{cases} \quad (5)$$

### 3.2 Hybrid Approach

The proposed method although it improves the accuracy of recommendations, is unable to provide recommendations to other users that do not have at least a number of co-rated items and a certain PCC similarity value. For this reason, a hybrid approach that can switch to PCC only if enough ratings are not available for the multi-level approach to provide recommendations. Algorithm 1 provides the hybrid approach.

**Algorithm 1:** A hybrid approach to recommendations

**Input:** User id /* the id of the user requesting the recommendations */
t /* m is the minimum pre-specified number of common ratings – Valid values are t1, t2, t3 and t4 */
y /* y is the minimum pre-specified similarity value derived from PCC */
**Output:** Recommendation method

```
Load k nearest neighbors for User id
For (int i=0; i<k.size; i++)
        If k.get(i, n) >= t && s >=y  /* n is the number of co-rated items */
                                        /* s is the similarity value derived from PCC */
                Then Load function defined in equation 5
        Else
                Load PCC
        End If
End For
```

## 4. Experimental Evaluation

In this section the experimental evaluation of our proposed method takes place and the results are based on five real datasets and a number of widely used metrics with different parameters. The evaluation took place on an Intel i3 2.13 GHz, 4GBs of RAM, running windows 8.1 and all the algorithms were implemented in the Java programming language.

### 4.1 Real datasets

We conducted our experiments on five real datasets in order to observe the results under different amounts of ratings and users. The five datasets are the MovieLens 100 thousand (Herlocker et al., 1999), MovieLens 1 million (Herlocker et al., 1999), Jester dataset 2 (Goldberg et al., 2001), Epinions (Massa & Avesani, 2007) and MovieTweetings (Dooms et al., 2013).

- **MovieLens 100 thousand.** MovieLens 100 thousand is a real dataset which contains 1682 movies, 943 users and 100.000 ratings. The data have been collected by the University of Minnesota and are associated with their online movie recommendation system. This particular dataset is one of the many that is publicly available from the University and has been widely used before for offline experimental evaluation of collaborative filtering recommender system performance. The data in the dataset are in the form [userid] [itemid] [rating]. All the rating values are in the scale 1 to 5.

- **MovieLens 1 million.** MovieLens 1m is a real dataset which contains 4000 movies, 6000 users and 1,000,000 ratings. The data have been collected by the University of Minnesota and are associated with their online movie recommendation system. This particular dataset is one of the many that is publicly available from the University and has been widely used before for offline experimental evaluation of collaborative filtering recommender system performance. The data in the dataset are in the form [userid] [itemid] [rating]. All the rating values are in the scale 1 to 5.

- **Jester.** This is a publicly available dataset developed at the University of California, Berkeley and has over 1.7 million ratings of 150 jokes from 59.132 users. The data from the dataset are associated to their online joke recommendation system. Like the two previous datasets, this one has been used widely for offline experimental evaluation of collaborative filtering recommender systems and the data are in the form [userid] [itemid] [rating]. All the rating values are in the scale -10 to 10. Moreover, from this dataset we have used a subset consisting of the first 1 million entries.

- **Epinions.** This is a publicly available dataset crawled from Epinions.com. It is a general product recommendation dataset. The data in the dataset are in the form [userid] [itemid] [rating]. All the rating values are in the scale 1 to 5. The dataset has 664,824 ratings from 49,290 users on 139,738 items.

- **MovieTweetings.** This is a publicly available dataset crawled from Twitter. It is a dataset consisted of movie ratings in the scale 0 to 10. The data in the dataset are in the form [userid] [itemid] [rating]. The dataset has 431,780 ratings from 39,363 users on 22,610 items.

### 4.2 Comparisons

We have made used the following collaborative filtering recommendation methods in the comparisons.

- **PCC.** This is a method that calculates the statistical correlation between the common ratings of two users in order to determine the similarity between them. The output will be between -1, which is the lowest, and 1, which is the highest possible value. This method is defined in equation 1.

- **WPCC.** This is a method that is based on PCC. The difference is that the algorithm considers a pre-defined number of common ratings between users. Moreover if the number of common ratings is not sufficient then it switches to classical PCC.

- **SPCC.** This is a method that is based on PCC. The difference is that the similarity value returned for users with a small number of common items is weaker. However, we have adjusted the function to produce a higher similarity to users with a smaller number of co-rated items.

The aforementioned methods that have been compared against our proposed method have different characteristics. Our proposed method is based on multiple levels, from top-to-bottom, with each of these levels having a number of constraints. The constraints provide a higher similarity value between users that have more common items and a PCC similarity value above a certain threshold, which is something that is not available in the other methods. However, it should be noted that if any of the constraints of the level is not satisfied then the similarity value between the users will be set to zero. Thus, the main weakness of the proposed algorithm is that in the case that enough ratings are not available, then none of the levels can be constructed and recommendations cannot be provided. In this case another recommendation algorithm, such as PCC, needs to be applied. Furthermore, Table 3 provides a comparison between the methods.

| **Method** | **Characteristics** |
|---|---|
| PCC | Statistical correlation between two users providing a similarity value from -1 to 1 |
| WPCC | Based on PCC, gives more weight to users with a pre-specified number of common ratings |
| SPCC | Extends PCC by providing a weaker similarity value between users depending on co-rated items |
| Proposed | Enchases the similarity value of users that belong to certain categories and ignores the rest |

**Table 3.** Comparison between methods

### 4.3 Measures

For the purpose of measuring the accuracy of the recommendations provided by collaborative filtering recommendations the widely accepted by the research community Mean Absolute Error (MAE) metric has been used (Herlocker et al., 2004; Shani & Gunawardana, 2011). MAE is defined in equation 6, where $pi$ is the predicted rating and $ri$ is the actual rating in the summation. This method is used for the computation of the deviation between the predicted ratings and the actual ratings. It should also be noted that lower values are better.

$$MAE = \frac{1}{n}\sum_{i=1}^{n}|pi - ri| \quad (6)$$

In information retrieval systems, such as recommender systems, there are metrics that can measure the quality of the top N recommendations (Herlocker et al., 2004; Shani & Gunawardana, 2011). These metrics are Precision and Recall. For these metrics, higher values are better. Equation 7 defines the Precision metric, which is the amount of relevant recommendations found in the retrieved set of recommendations and Recall is the amount of relevant recommendations that have been retrieved successfully. Equation 8 contains the definition of Recall metric.

$$precision = \frac{Correctly\ recommended\ items}{Total\ recommender\ items} \quad (7)$$

$$recall = \frac{Correctly\ recommended\ items}{Relevant\ items} \quad (8)$$

### 4.4 Settings

For the below experimental evaluation process, the equation number 5 is used, which defines the proposed method. The following settings have been used:

- **Values** $x$ **and** $y$ of the proposed method. $x$ is a pre-defined value that is added to the similarity value returned from PCC and has been set 0.50 for the first(top) level, 0.375 for the second, 0.25 for the third and 0.125 for the last level. $y$ is the similarity value returned from PCC and has been set to greater than or equal to 0.33.

- **WPCC** recommendation method settings. For the MovieLens 100k and 1m datasets this number has been set to 50. For the Jester, Epinions and MovieTweetings datasets has been set to 5.

- **MAE.** For the MAE evaluation, all the datasets have been split into two parts. An 80% randomly selected part that has been used for training and the remaining 20% used for testing. In the experimental results in section 4.5, $k$ is the number of neighbors.

- **Precision and Recall.** For these metrics two experiments were conducted. The first one uses a five user neighborhood and asks for the top 5 recommendations. The second one uses a ten user neighborhood and asks for 10 recommendations. In the experimental results in section 4.5, $k$ is the number of neighbors and *r* is the number of recommendations requested. We must notice that, researchers state that measuring Recall is not practical in recommender systems and that Precision is more important in top N recommendations and the output values depend on the number of the recommendations requested and the size of the nearest neighborhood (Herlocker et al., 2004; Liu et al., 2014).

- **t1, t2, t3 and t4.** The values for these variables have been set to t1=50 t2=20 t3=10 t4=5. According to function 5 the number of co-rated items for level 1 must be equal or greater than 50, level 2 is between 49 and 20, level 3 is between 19 and 10 and level 4 is between 9 and 5.

### 4.5 Experimental results

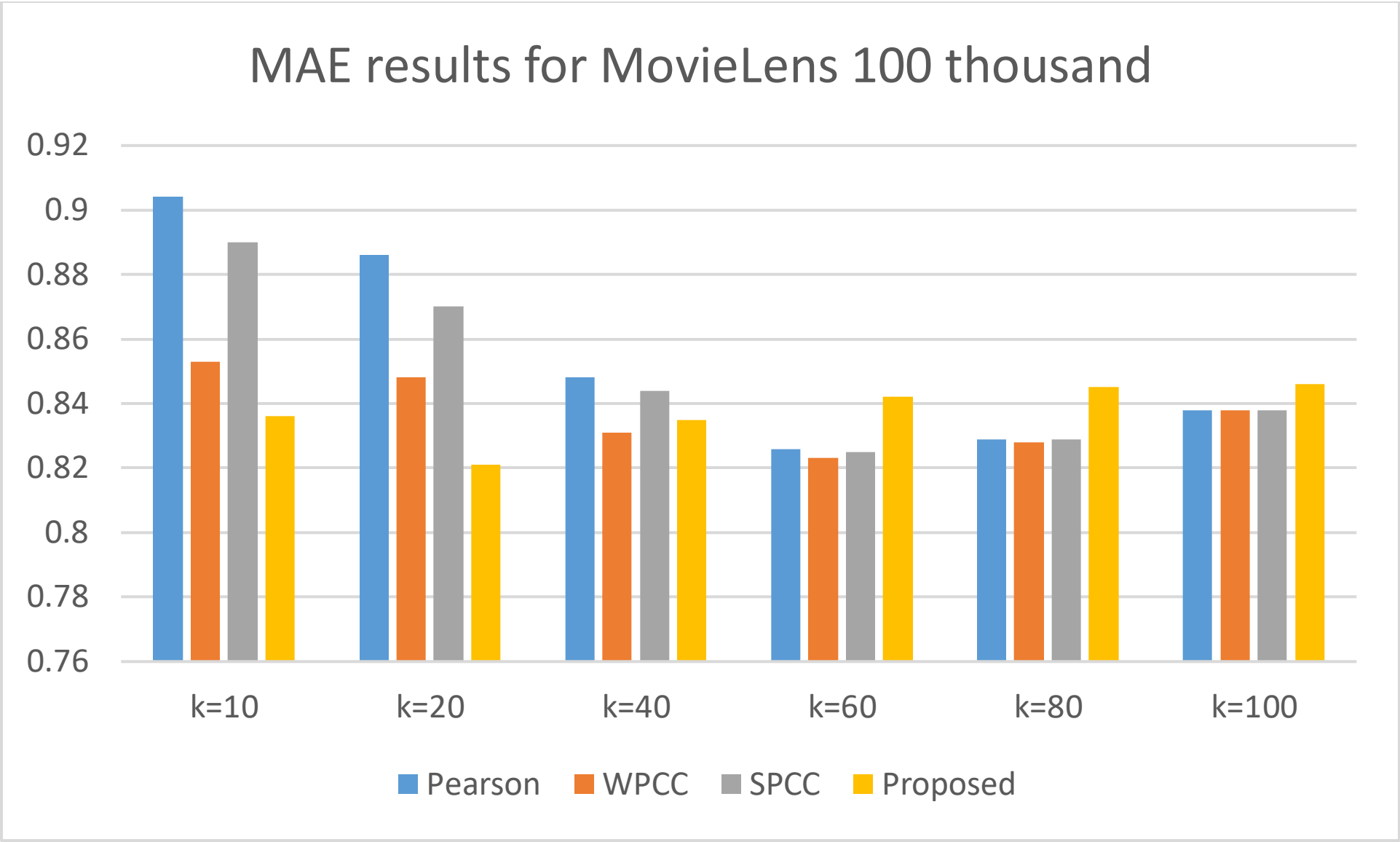

**Figure 1.** MAE results for the MovieLens 100.000 dataset

The MAE results for the MovieLens 100k dataset are shown in figure 1. It is shown that when the neighborhood is small, up to 20 users, our proposed method outperforms all the other recommendation methods. When the number of neighbors is getting higher, i.e 40 or more, we can see that our proposed method becomes less effective.

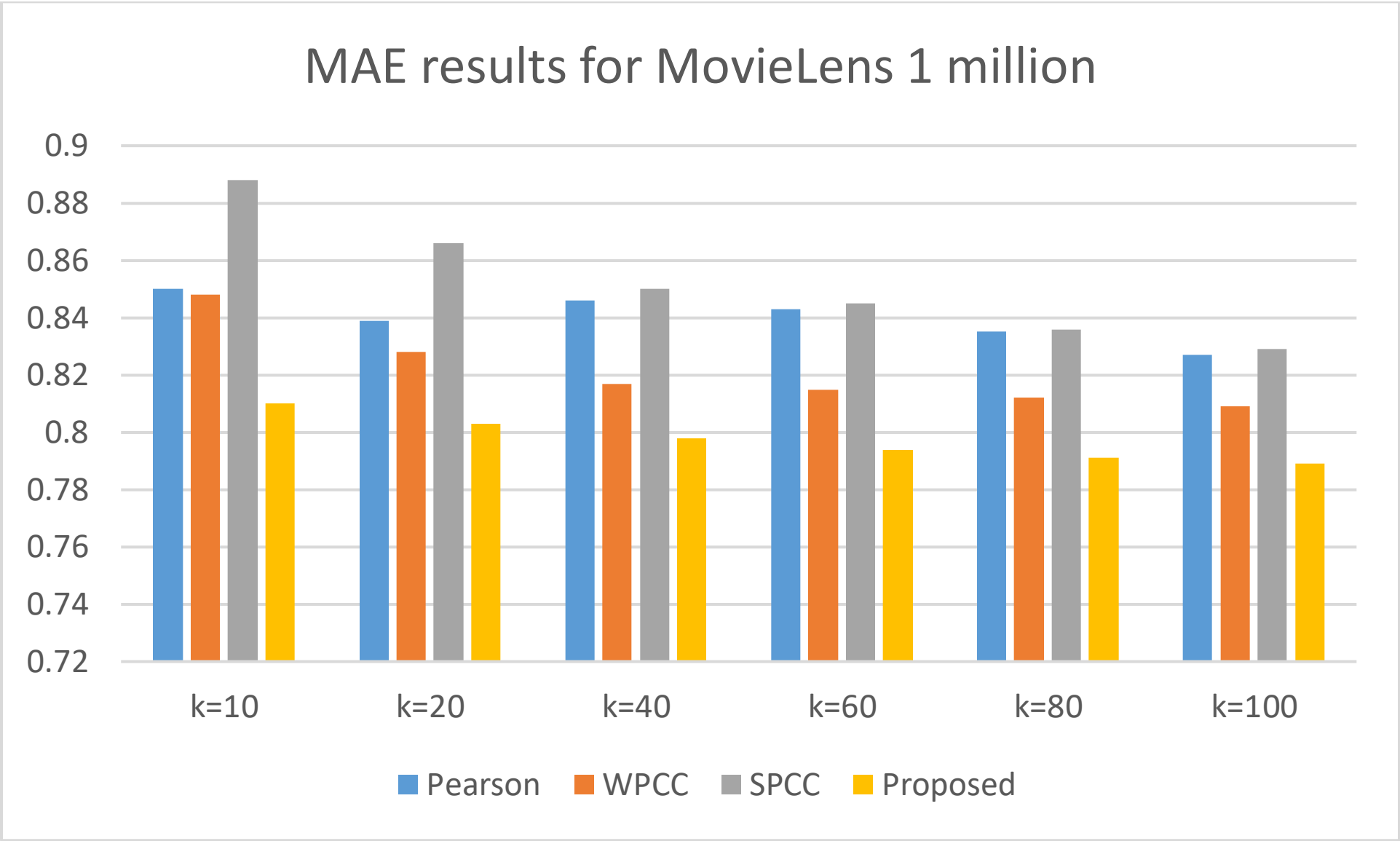

**Figure 2.** MAE results for the MovieLens 1.000.000 dataset

The MAE results for the MovieLens 1m dataset are shown in figure 2. It is shown that our proposed method ourperforms the other methods. Besides that, we can see that as the number of neighbors is getting higher our proposed method becomes more effective. We can see that the larger the neighbrhood grows the results become better for all methods. This is due to the fact that a larger number of users is available to his dataset compared to the 100k one.

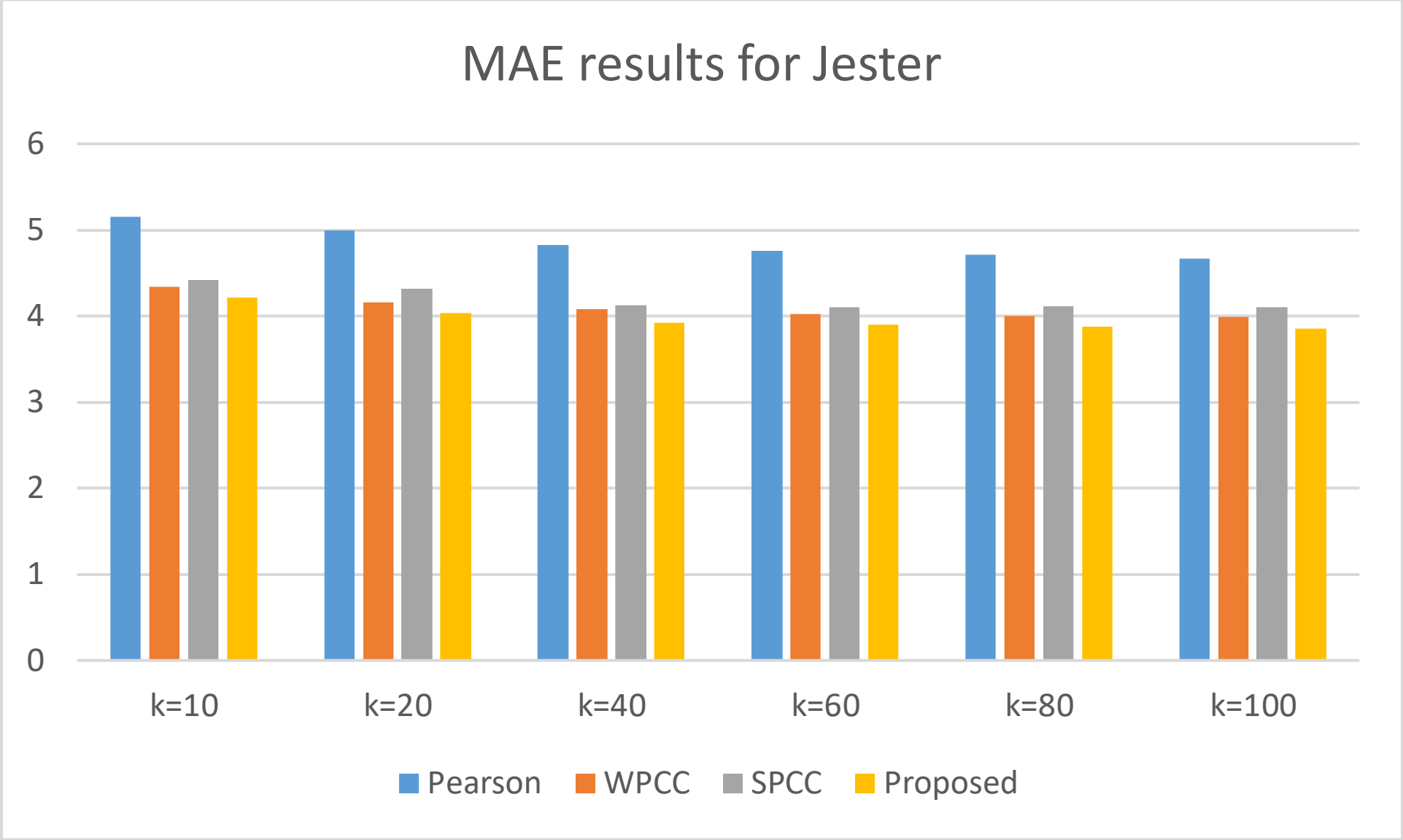

**Figure 3.** MAE results for the Jester dataset

The MAE results for the Jester dataset are shown in figure 3. It is shown that our proposed method ourperforms the other methods. However in all cases (although our method achieves a lower MAE) there aren't any significant changes when the size of the neighborhood grows. It is also remarkable that all the alternative recommendation methods have a noteworthy difference from PCC. Moreover, the results are almost identical when the neighborhood grows, since the number of users in the dataset is very small.

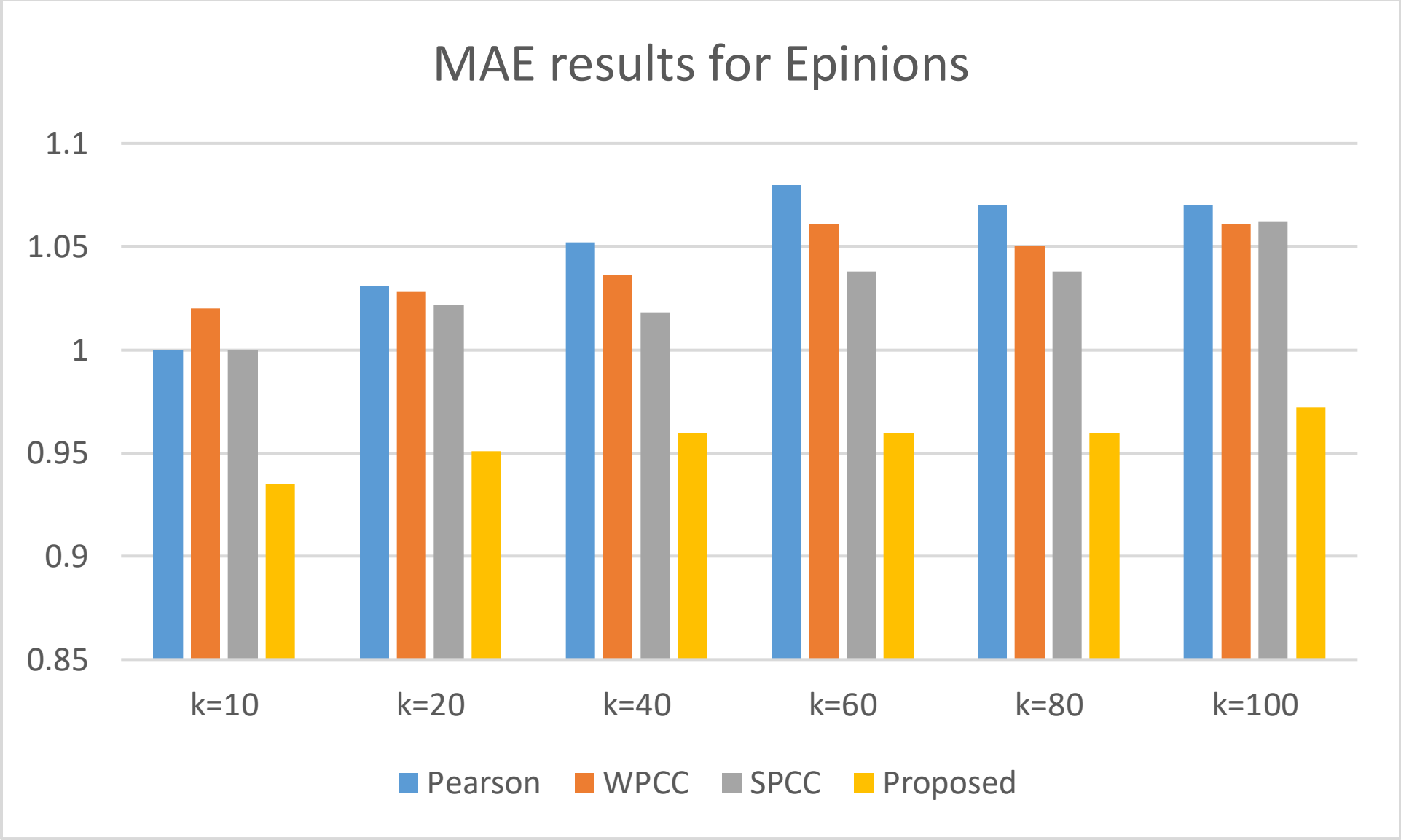

**Figure 4.** MAE results for the Epinions dataset

The MAE results for the Epinions dataset are shown in figure 4. It is shown that our proposed method ourperforms the other methods. However in all cases there aren't any significant changes between our method and the others when the size of the neighborhood grows.

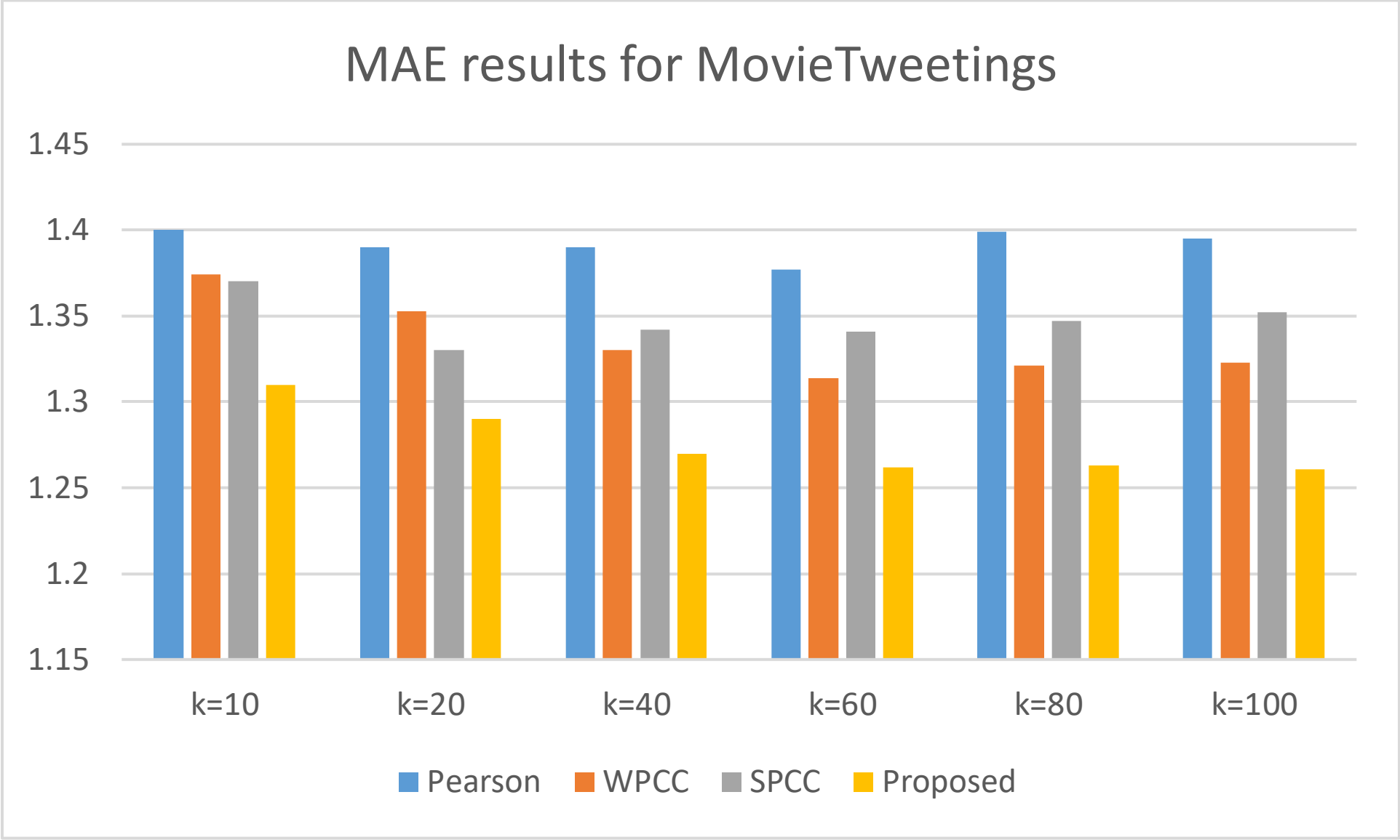

**Figure 5.** MAE results for the MovieTweetings dataset

The MAE results for the MovieTweetings dataset are shown in figure 5. It is shown that our proposed method ourperforms the other methods. Besides that, we can see that as the number of neighbors is getting higher our proposed method becomes more effective.

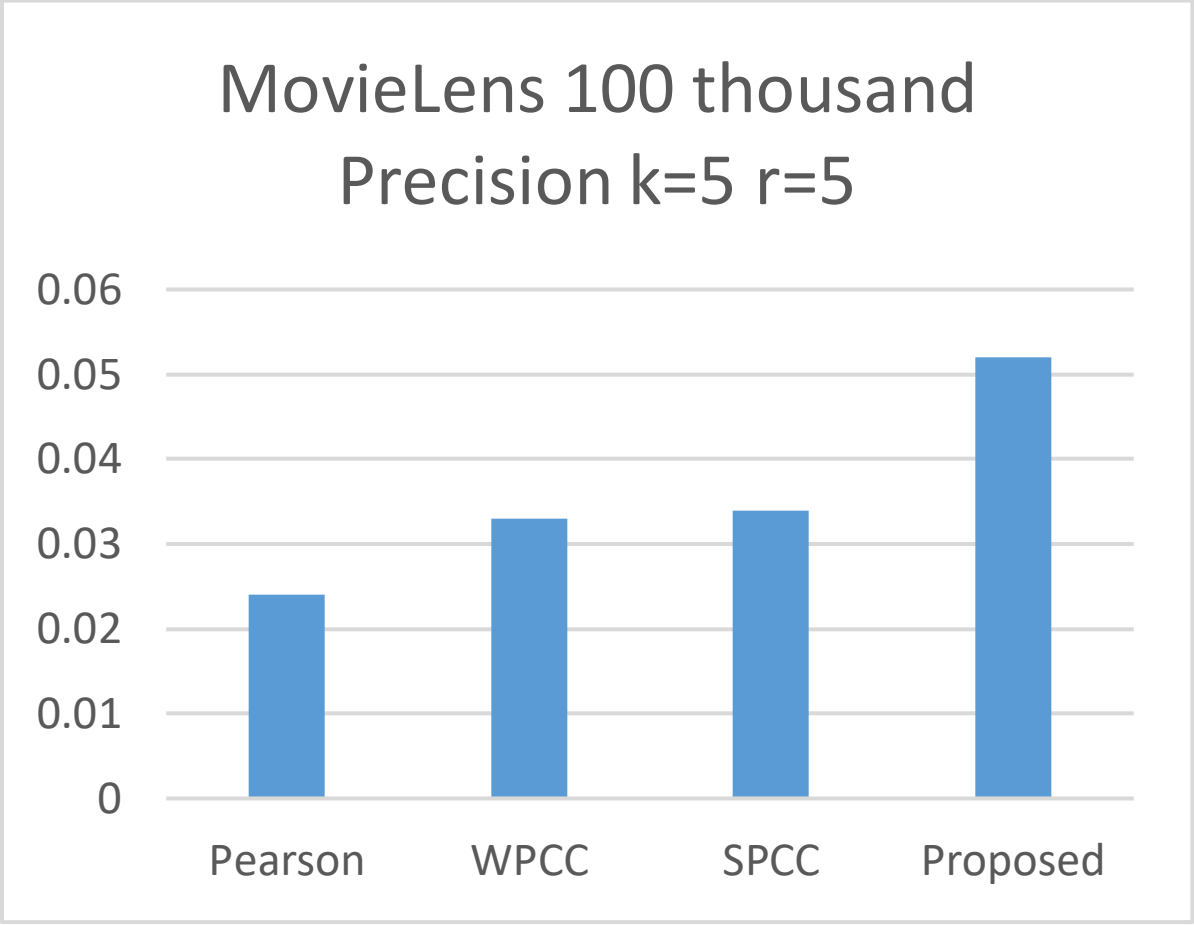

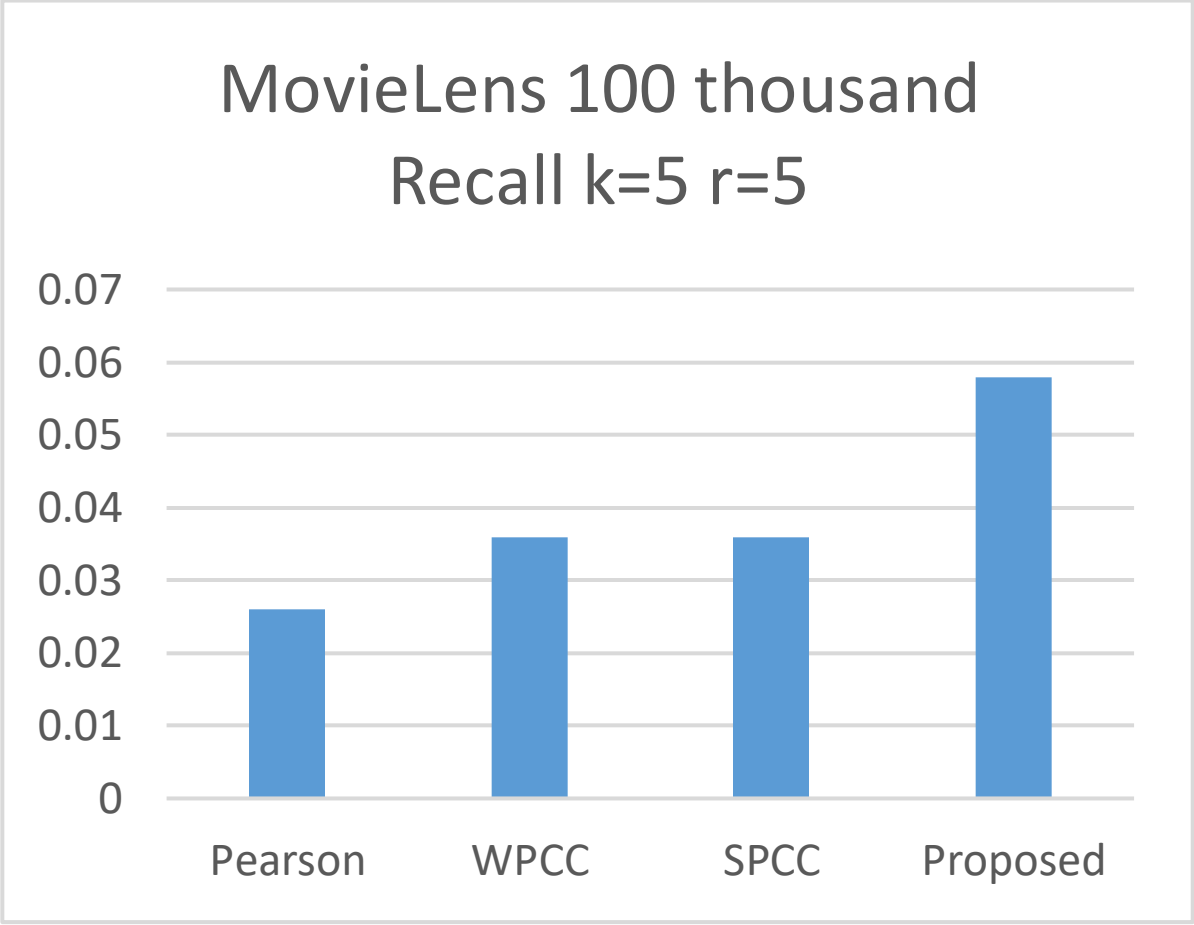

(a) (b)

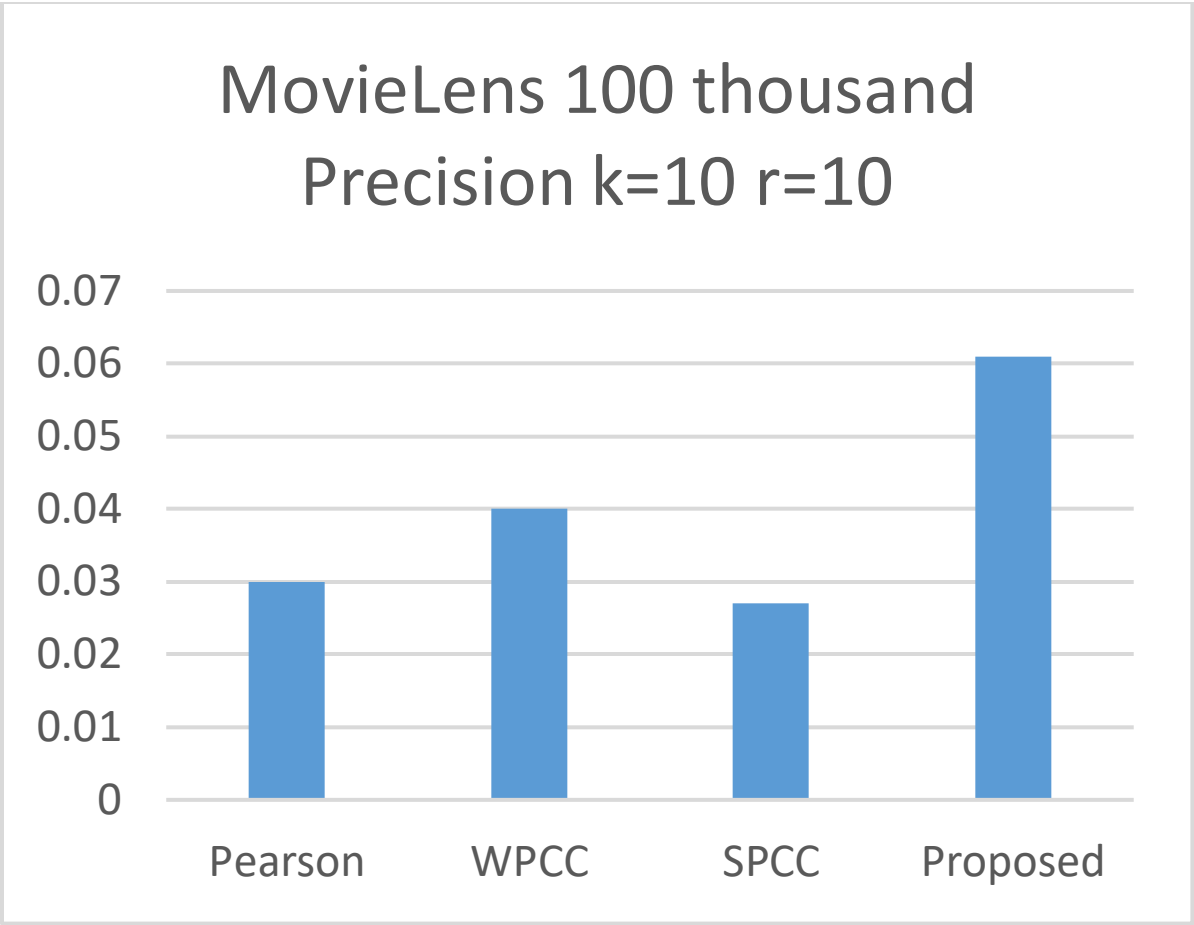

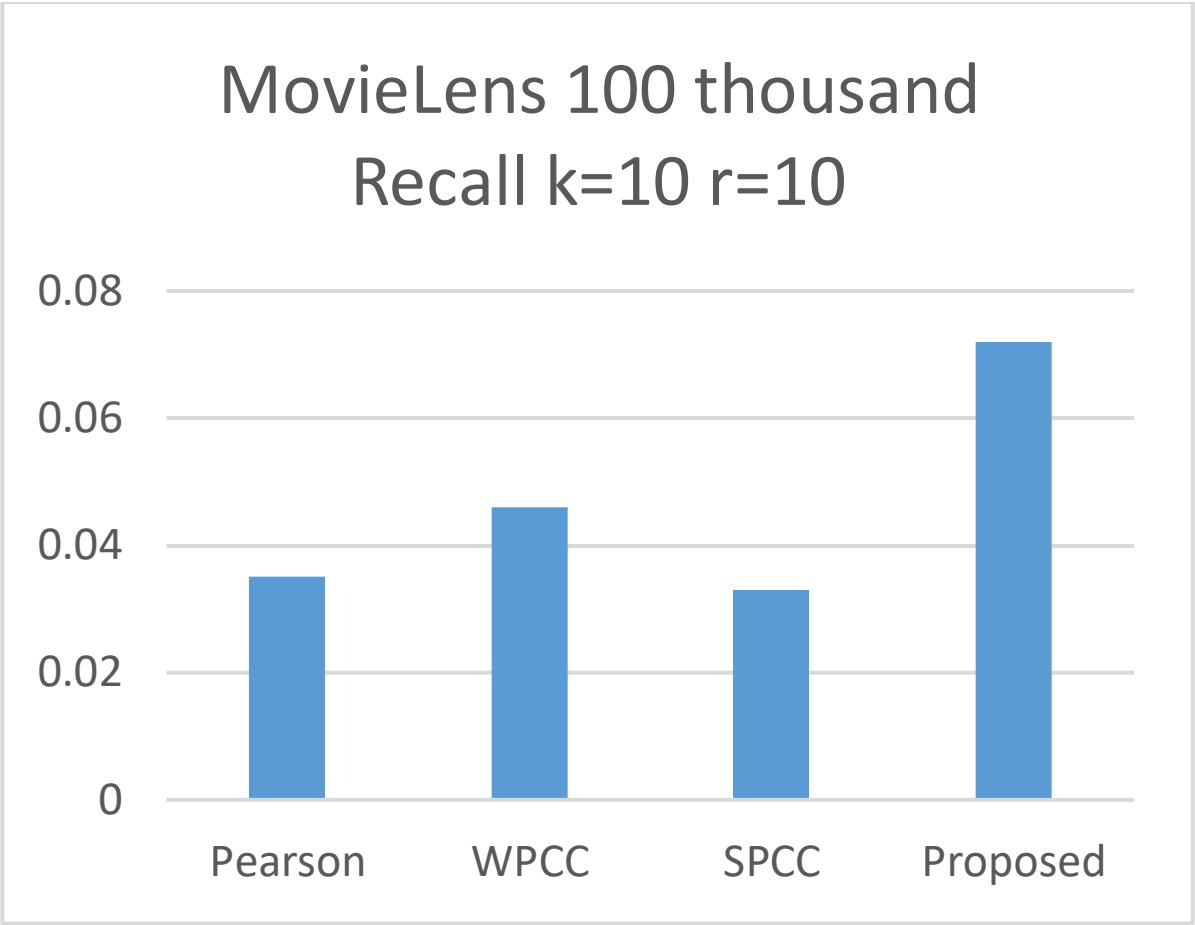

(c) (d)

**Figure 6.** Precision and Recall results for the MovieLens 100 thousand dataset

In figure 6 we can see the results obtained by using the Precision and Recall metrics using the MovieLens 100k dataset. Sub figure 6,a represents the Precision results for 5 neighbors and 5 recommendations. Sub figure 6,b represents the Recall results for 5 neighbors and 5 recommendations. Sub figure 6,c represents the Precision results for 10 neighbors and 10 recommendations. Sub figure 6,d represents the Recall results for 10 neighbors and 10 recommendations. It is shown that as the neighborhood grows larger, the quality of the top N recommendations becomes better.

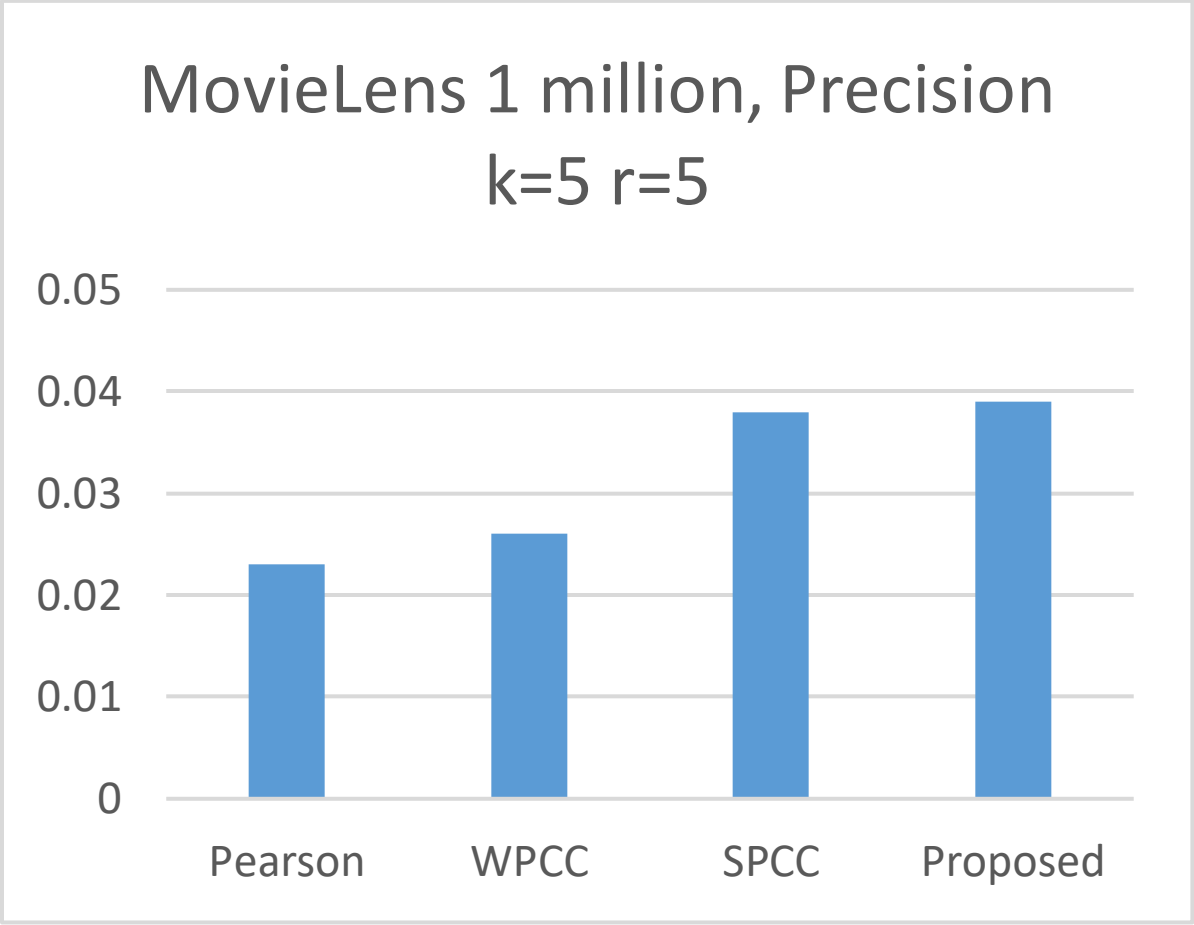

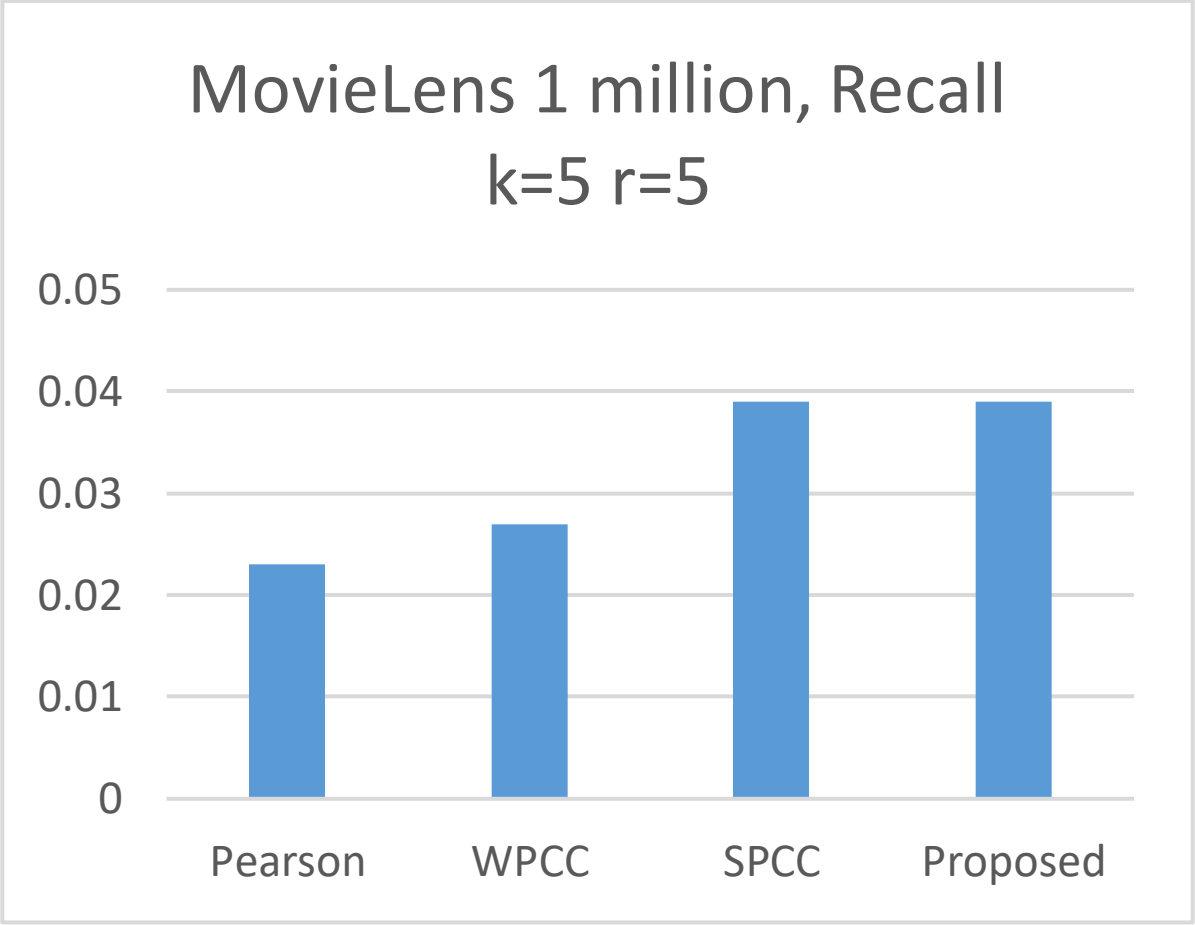

(a) (b)

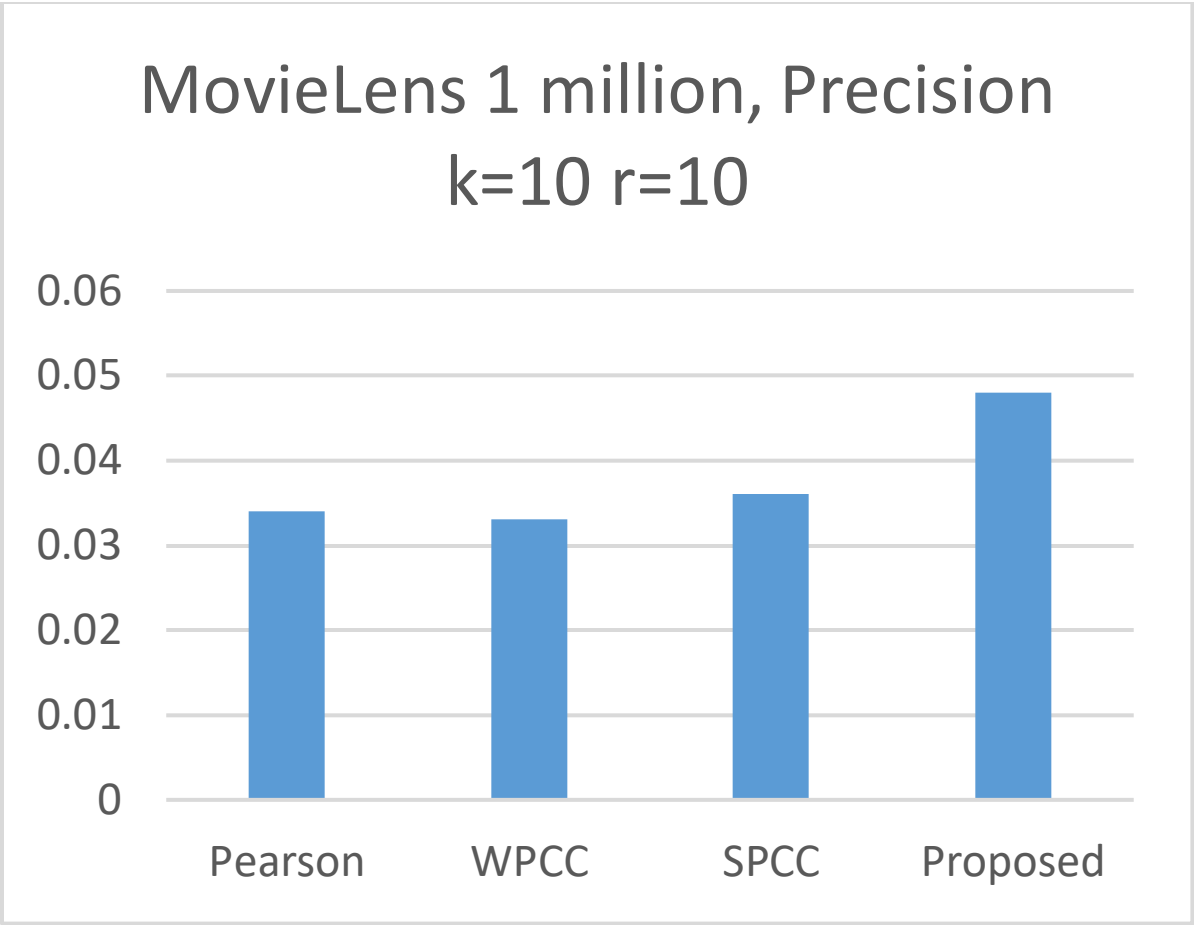

(c)

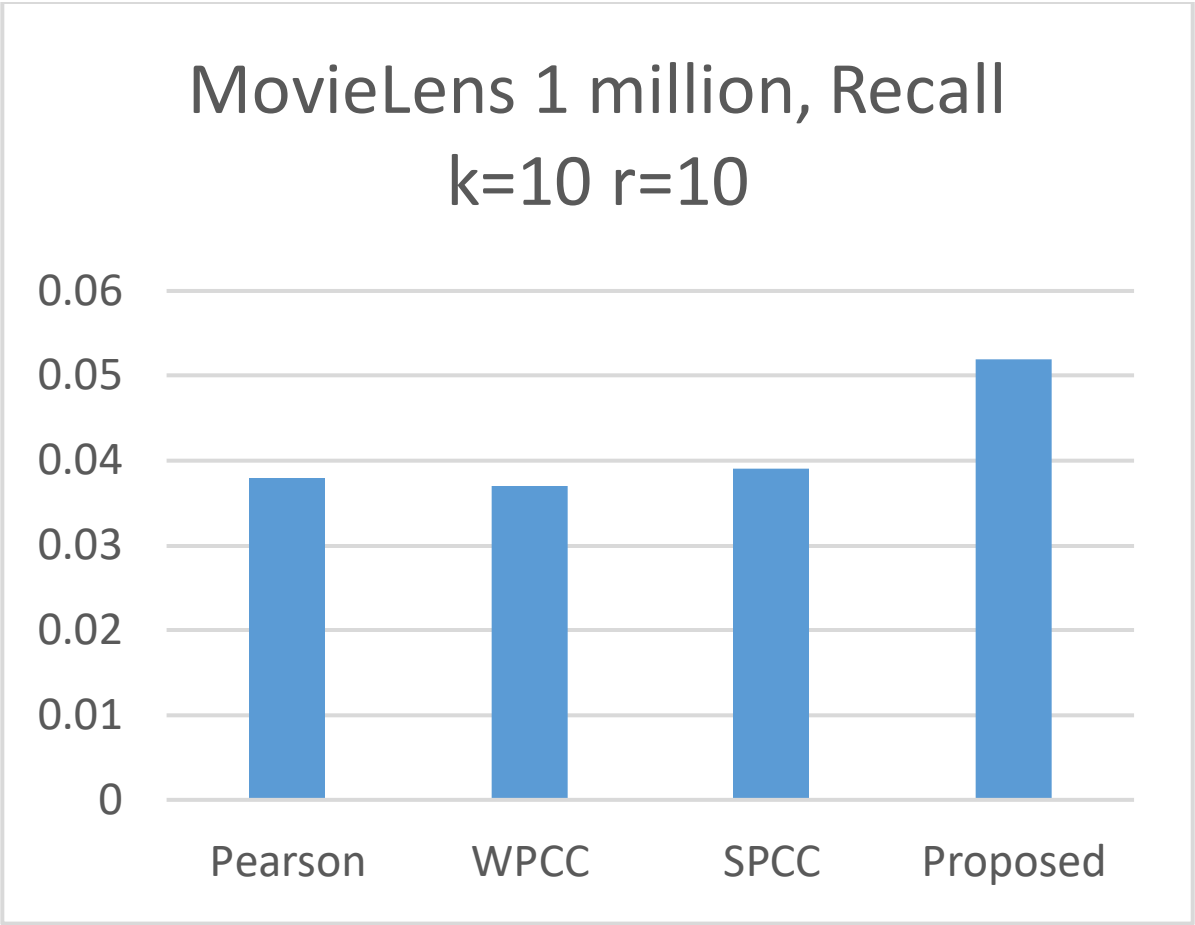

(d)

**Figure 7.** Precision and Recall results for the MovieLens 1 million dataset

In figure 7 we can see the results obtained by using the Precision and Recall metrics using the MovieLens 1m dataset. Sub figure 7,a represents the Precision results for 5 neighbors and 5 recommendations. Sub figure 7,b represents the Recall results for 5 neighbors and 5 recommendations. Sub figure 7,c represents the Precision results for 10 neighbors and 10 recommendations. Sub figure 7,d represents the Recall results for 10 neighbors and 10 recommendations. It is shown that as the neighborhood grows larger, the quality of the top N recommendations becomes better.

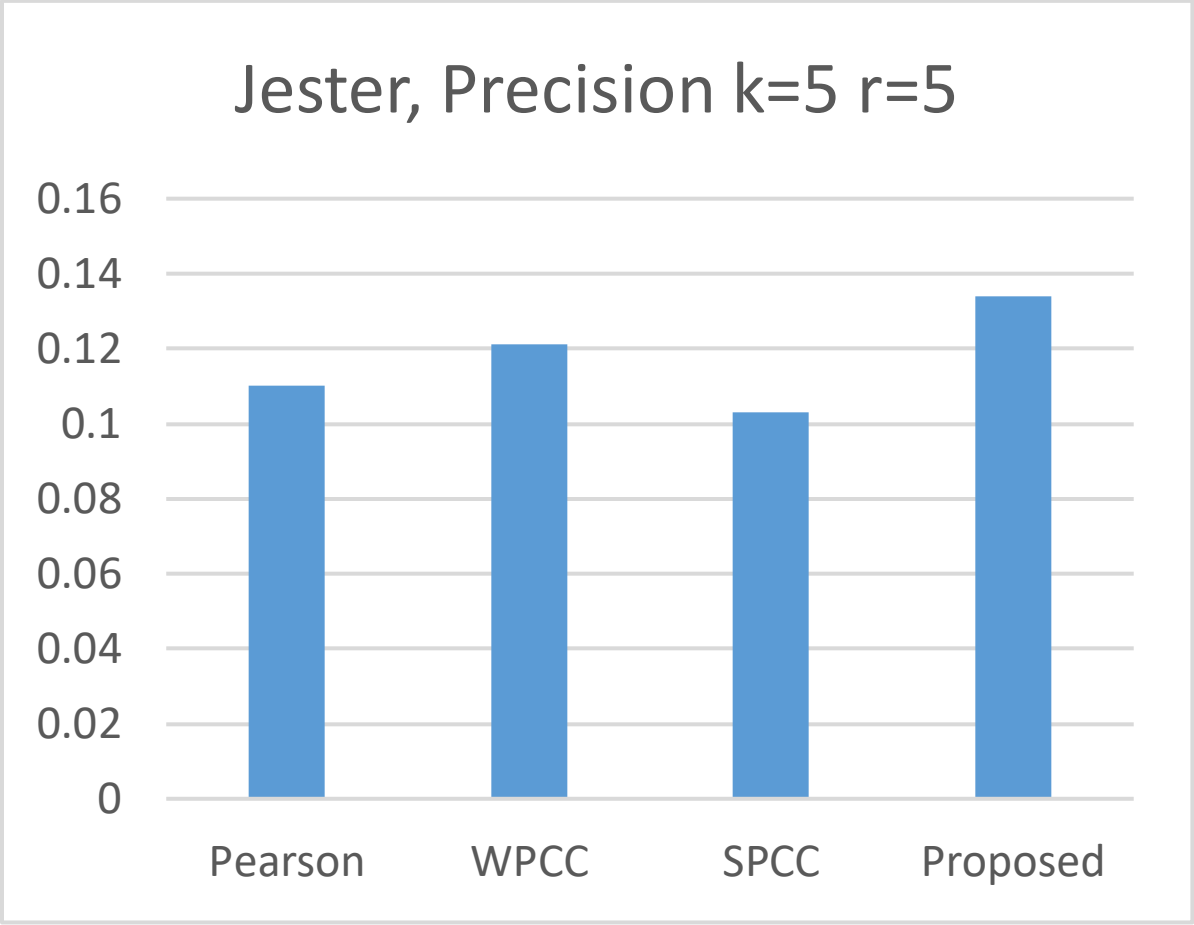

(a)

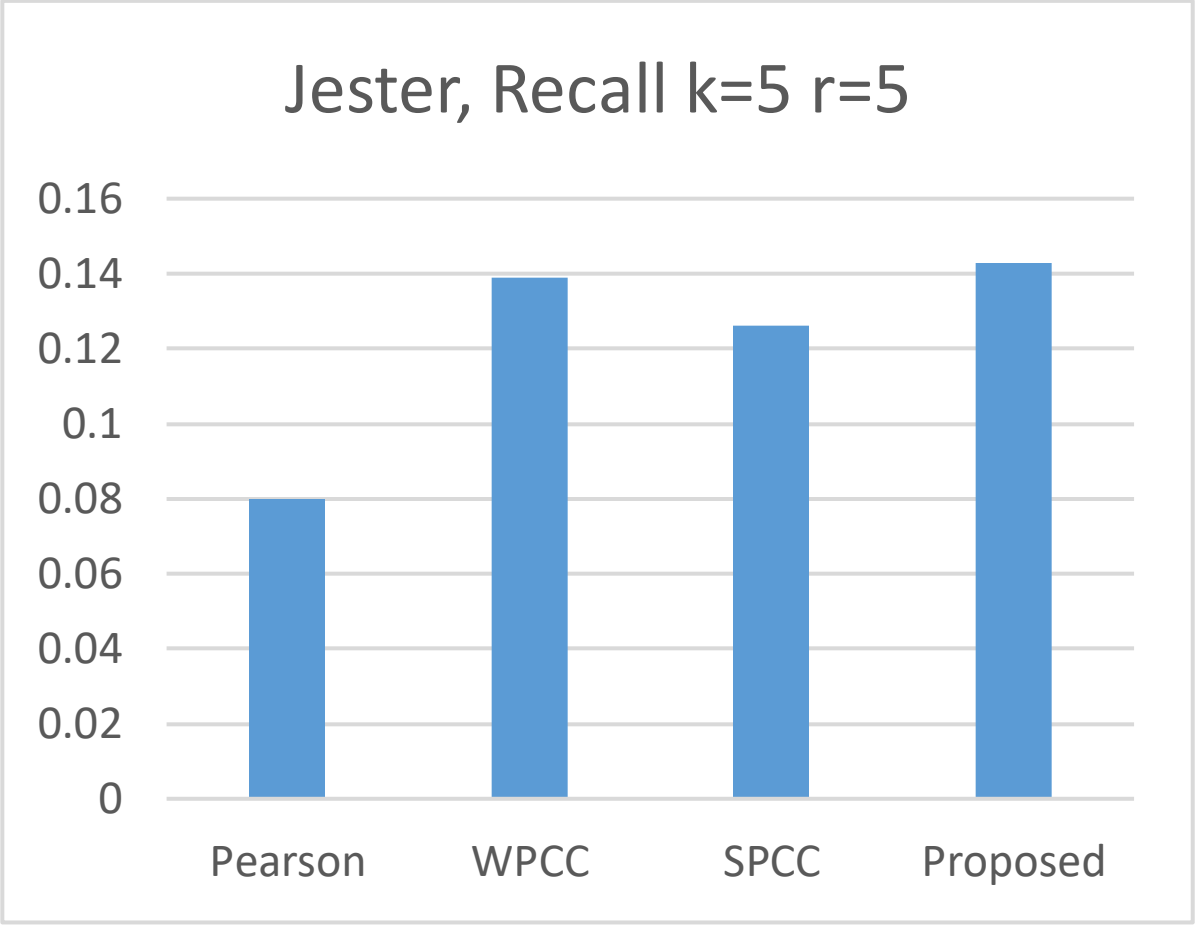

(b)

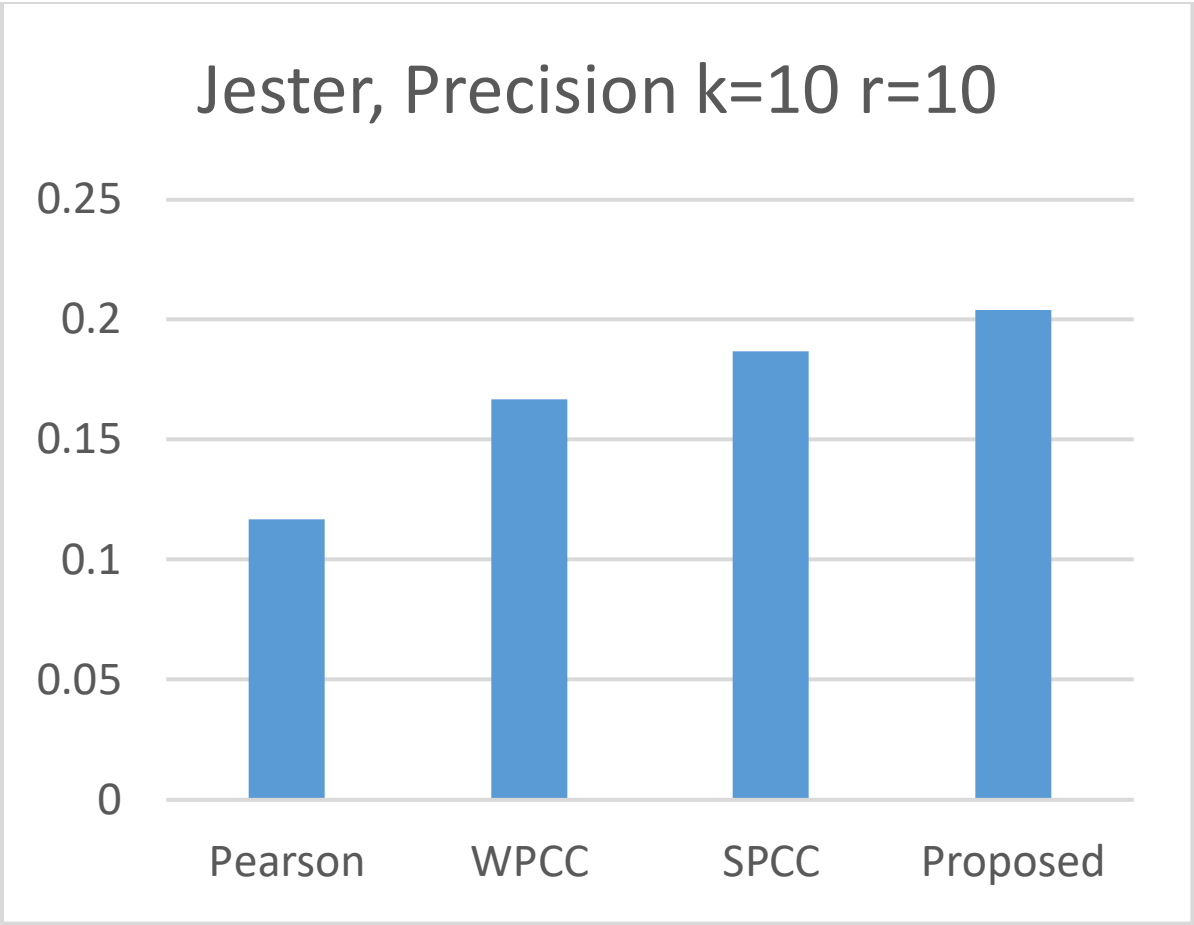

(c)

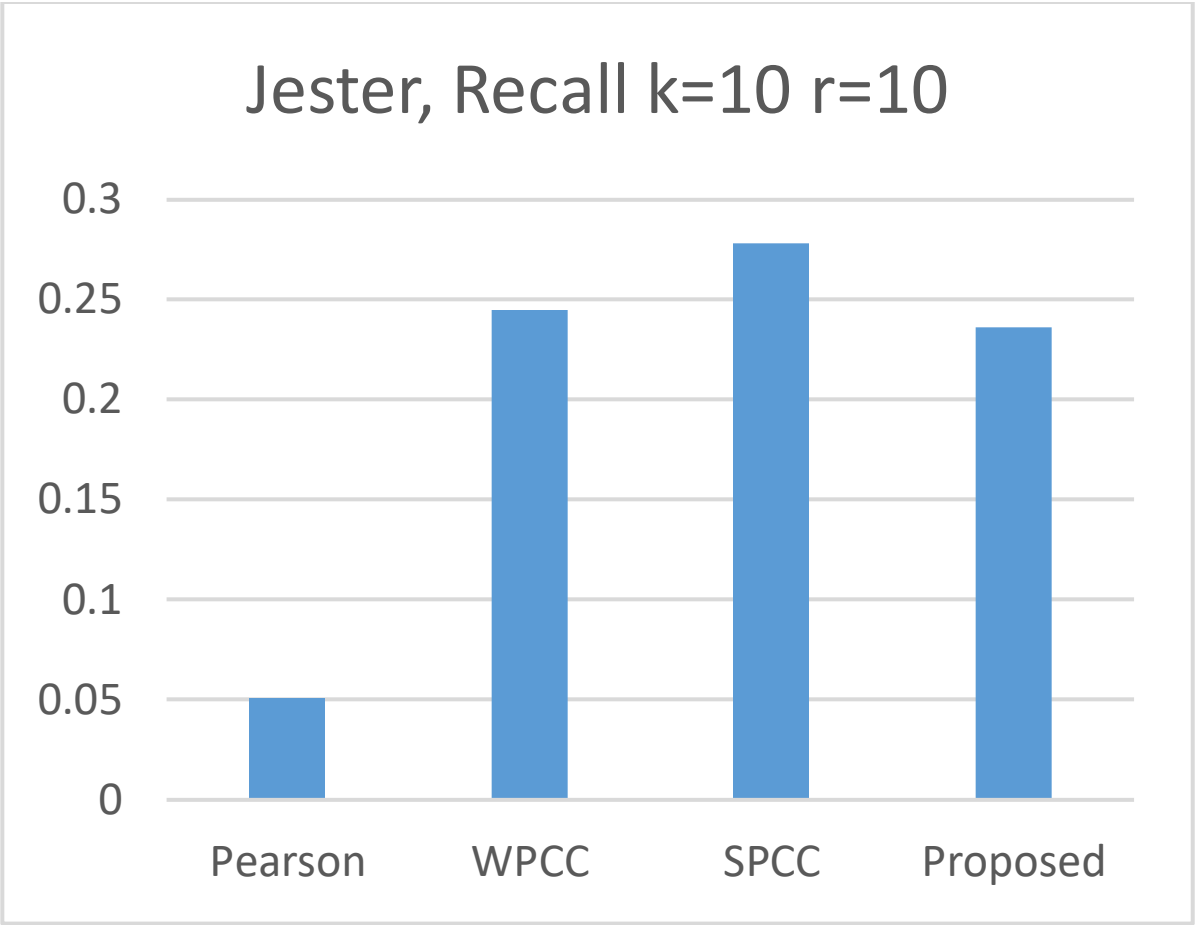

(d)

**Figure 8.** Precision and Recall results for the Jester dataset

In figure 8 we can see the results obtained by using the Precision and Recall metrics using the Jester dataset. Although our method provides good recommendations, when the neighborhood grows the quality of all the alternatives, except from PCC, is pretty much the same. This is due to the fact that the dataset is comprised with a small number of users. Sub figure 8,a represents the Precision results for 5 neighbors and 5 recommendations. Sub figure 8,b represents the Recall results for 5 neighbors and 5 recommendations. Sub figure 8,c represents the Precision results for 10 neighbors and 10 recommendations. Sub figure 8,d represents the Recall results for 10 neighbors and 10 recommendations

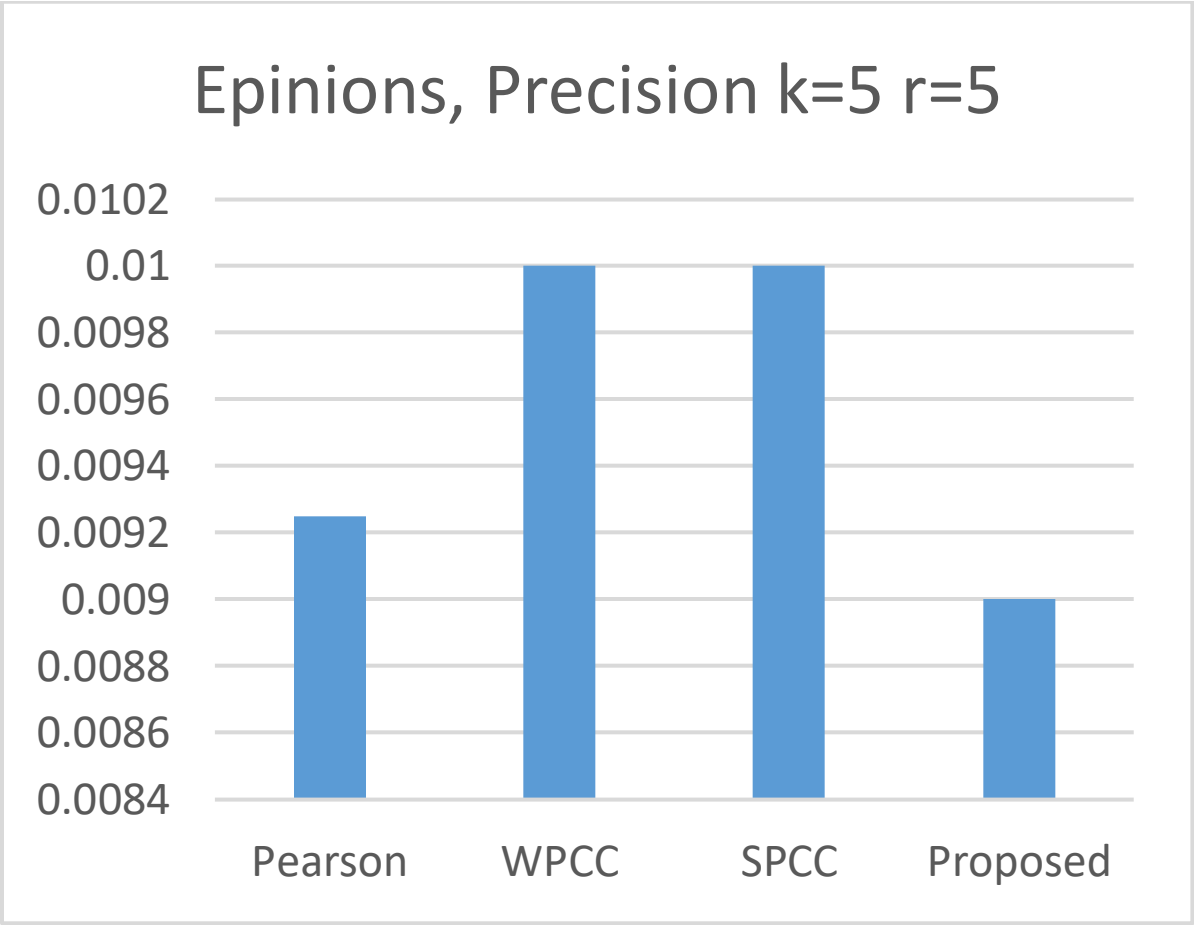

(a)

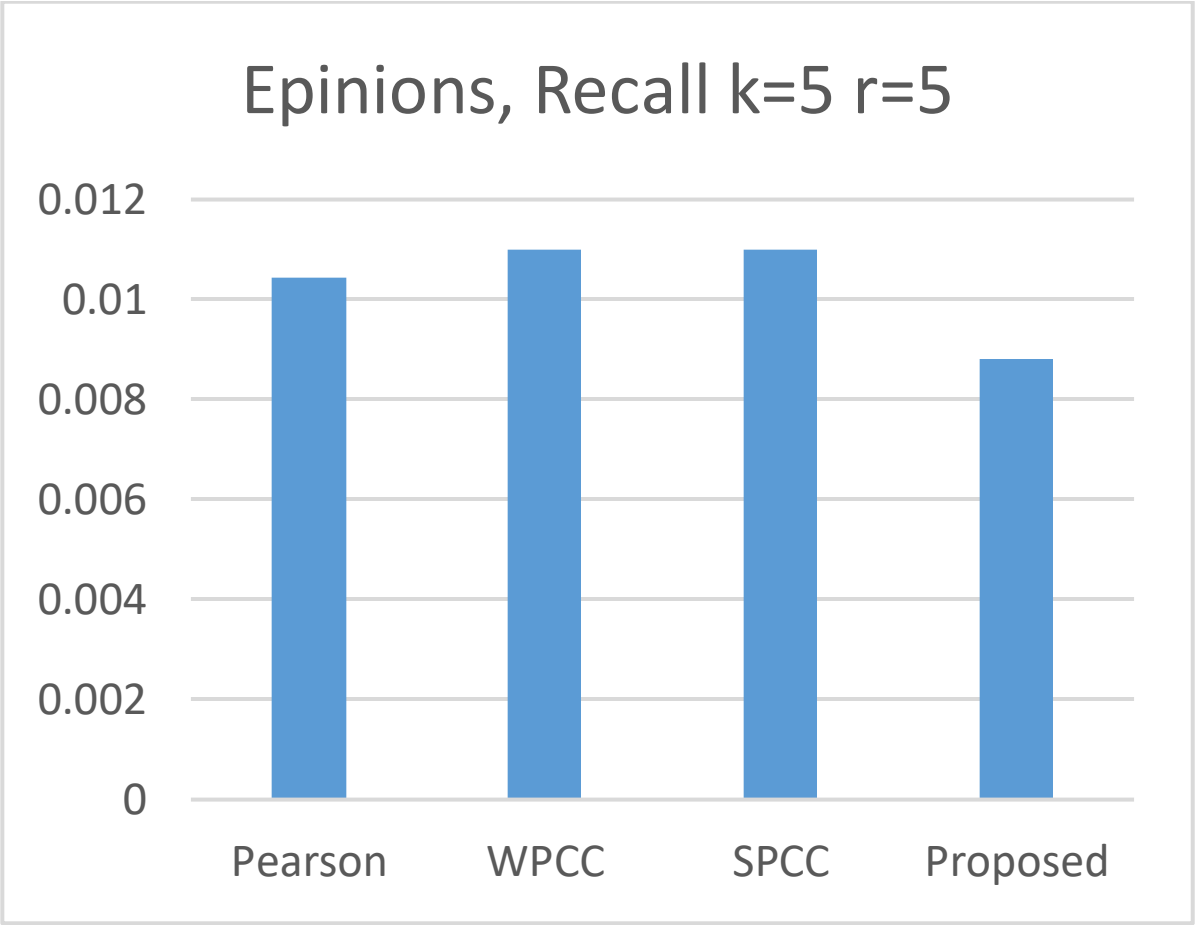

(b)

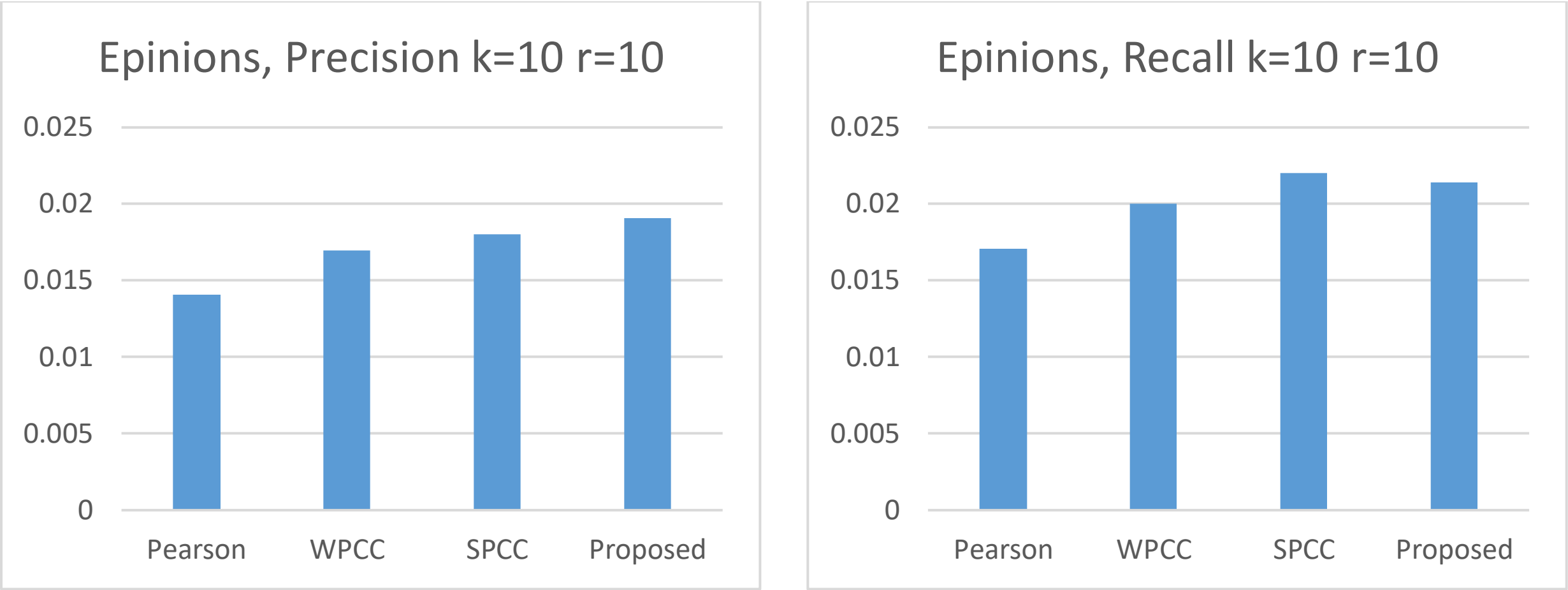

(c) (d)

**Figure 9.** Precision and Recall results for the Epinions dataset

In figure 9 we can see the results obtained by using the Precision and Recall metrics using the Epinions dataset. Sub figure 9,a represents the Precision results for 5 neighbors and 5 recommendations. Sub figure 9,b represents the Recall results for 5 neighbors and 5 recommendations. Sub figure 9,c represents the Precision results for 10 neighbors and 10 recommendations. Sub figure 9,d represents the Recall results for 10 neighbors and 10 recommendations. It is shown that as the neighborhood grows larger, the quality of the top N recommendations becomes better.

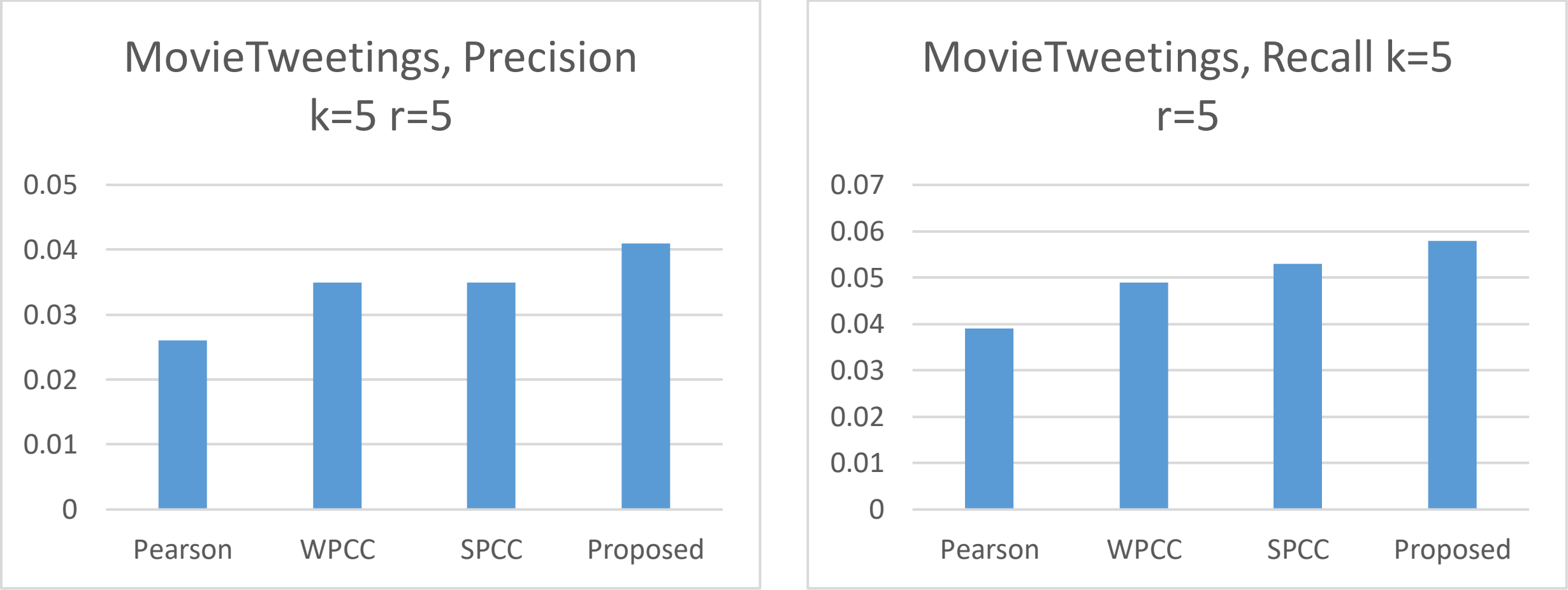

(a) (b)

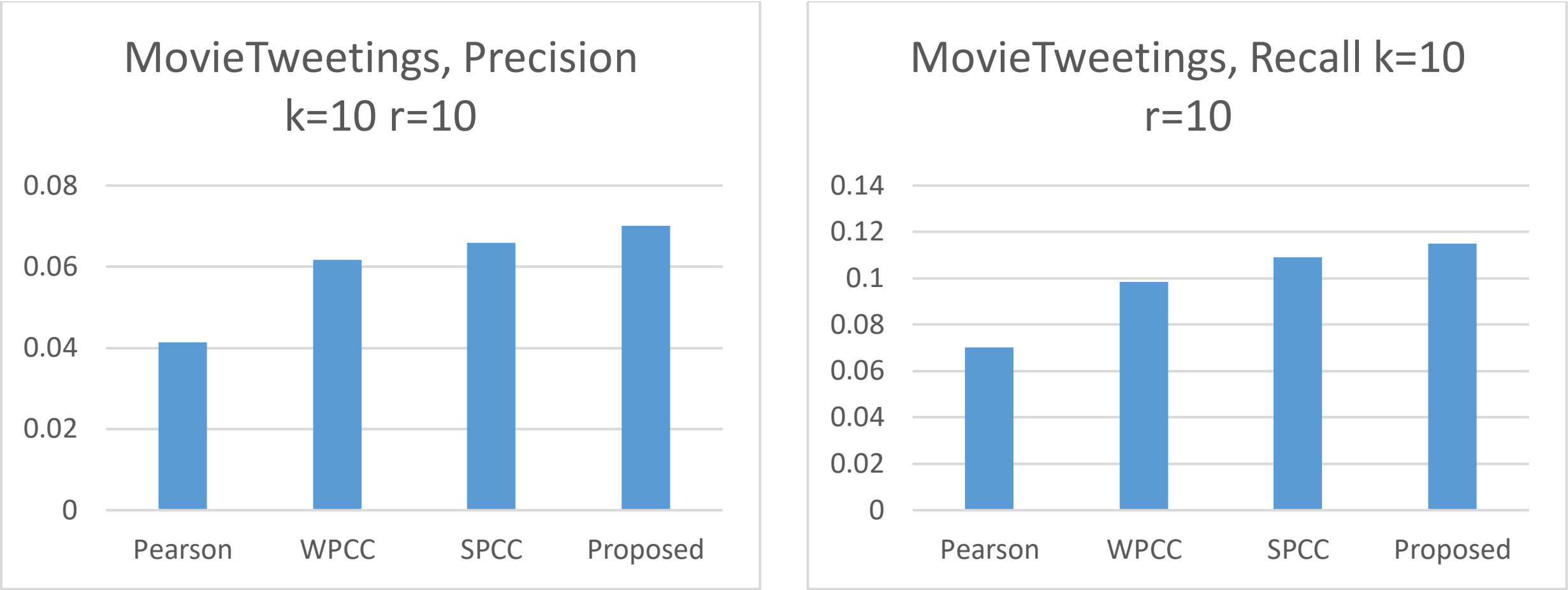

(c) (d)

**Figure 10.** Precision and Recall results for the MovieTweetings dataset

In figure 10 we can see the results obtained by using the Precision and Recall metrics using the MovieTweetings dataset. Sub figure 10,a represents the Precision results for 5 neighbors and 5 recommendations. Sub figure 10,b represents the Recall results for 5 neighbors and 5 recommendations. Sub figure 10,c represents the Precision results for 10 neighbors and 10 recommendations. Sub figure 10,d represents the Recall results for 10 neighbors and 10 recommendations. It is shown that our proposed method produces results of better quality in all cases.

## 5. Conclusions and Future Work

Although collaborative recommender systems have matured as a research area there are still issues related to their accuracy. On the other hand, people want to use an intelligent system to assist them in the decision making process in various online environments such as university and commerce domains among others. Furthermore, a recommender system needs to supply as accurate recommendations as possible, since its job is to substitute a human expert in a particular area. Thus, we have proposed a multi-level recommendation method that improves the accuracy of collaborative filtering. Aiming, at providing a better experience for decision making of users in online environments where the use of such systems is essential.

The main achievement of our method is to provide recommendations of higher accuracy this comes to a cost. There are cases that the similarity between two users is returned as zero and a set of recommendations cannot be provided. This is due to the fact that the similarity of uses is enchased only if they can be assigned to a certain level, after satisfying the required constraints, otherwise zero value is returned between them. Therefore, a different similarity measure or recommendation method needs to be uses until enough ratings are submitted to the database in order to make the proposed-multi-level approach usable.

Our future work will concentrate on dynamically enchasing or reducing the similarity value between users if certain conditions hold. As a result, the substitution of the fixed number of levels will be necessary and the conditions that will affect the similarity will change according to the available ratings, the common ratings and the similarity value between users at the given time point. Furthermore, the proposed method should be able to be adapted in context-aware collaborative filtering domains and this being another future work research part. Additionally, good explanations play a significant part in the recommendations process and a newly applied method needs to provide explanations of high quality to increase the confidence of users. Recommenders being a sub-field of expert systems need to give users an expert explanation on why a certain recommendation was provided. Moreover, the detection and removal of noisy users will be investigated. In collaborative recommendations, particularly in e-commerce, various users tend to give only negative or only positive ratings in order to affect the revenue of particular items or services. These types of users should be detected and removed.

## References

Ahn, H. J. (2008). A new similarity measure for collaborative filtering to alleviate the new user cold-starting problem. *Information Sciences*, *178*(1), 37-51.

Anand, D., & Bharadwaj, K. K. (2011). Utilizing various sparsity measures for enhancing accuracy of collaborative recommender systems based on local and global similarities. *Expert systems with applications*, *38*(5), 5101-5109.

Bobadilla, J., Ortega, F., Hernando, A., & Bernal, J. (2012a). A collaborative filtering approach to mitigate the new user cold start problem. *Knowledge-Based Systems*, *26*, 225-238.

Bobadilla, J., Ortega, F., & Hernando, A. (2012b). A collaborative filtering similarity measure based on singularities. *Information Processing & Management*, *48*(2), 204-217.

Bobadilla, J., Ortega, F., Hernando, A., & Gutiérrez, A. (2013). Recommender systems survey. *Knowledge-Based Systems*, *46*, 109-132.

Burke, R. (2002). Hybrid recommender systems: Survey and experiments. *User modeling and user-adapted interaction*, *12*(4), 331-370.

Burke, R. (2007). Hybrid web recommender systems. In *The adaptive web* (pp. 377-408). Springer Berlin Heidelberg.

Cacheda, F., Carneiro, V., Fernández, D., & Formoso, V. (2011). Comparison of collaborative filtering algorithms: Limitations of current techniques and proposals for scalable, high-performance recommender systems. *ACM Transactions on the Web (TWEB)*, *5*(1), 2.

Dooms, S., De Pessemier, T., & Martens, L. (2013). Movietweetings: a movie rating dataset collected from twitter. In Workshop on Crowdsourcing and human computation for recommender systems, CrowdRec at RecSys (Vol. 2013, p. 43).

Ekstrand, M. D., Riedl, J. T., & Konstan, J. A. (2011). Collaborative filtering recommender systems. *Foundations and Trends in Human-Computer Interaction*, *4*(2), 81-173.

Fang, H., Guo, G., & Zhang, J. (2015). Multi-faceted trust and distrust prediction for recommender systems. *Decision Support Systems*, *71*, 37-47.

Goldberg, K., Roeder, T., Gupta, D., & Perkins, C. (2001). Eigentaste: A constant time collaborative filtering algorithm. Information Retrieval, 4(2), 133-151.

Herlocker, J. L., Konstan, J. A., Borchers, A., & Riedl, J. (1999, August). An algorithmic framework for performing collaborative filtering. In *Proceedings of the 22nd annual international ACM SIGIR conference on Research and development in information retrieval* (pp. 230-237). ACM.

Herlocker, J. L., Konstan, J. A., Terveen, L. G., & Riedl, J. T. (2004). Evaluating collaborative filtering recommender systems. *ACM Transactions on Information Systems (TOIS)*, *22*(1), 5-53.

Jamali, M., & Ester, M. (2009, June). Trustwalker: a random walk model for combining trust-based and item-based recommendation. In *Proceedings of the 15th ACM SIGKDD international conference on Knowledge discovery and data mining* (pp. 397-406). ACM.

Jannach, D., Zanker, M., Felfernig, A., & Friedrich, G. (2010). *Recommender systems: an introduction*. Cambridge University Press.

Konstan, J. A., & Riedl, J. (2012). Recommender systems: from algorithms to user experience. *User Modeling and User-Adapted Interaction, 22*(1-2), 101-123.

Koutrika, G., Bercovitz, B., & Garcia-Molina, H. (2009, June). FlexRecs: expressing and combining flexible recommendations. In *Proceedings of the 2009 ACM SIGMOD International Conference on Management of data* (pp. 745-758). ACM.

Liu, H., Hu, Z., Mian, A., Tian, H., & Zhu, X. (2014). A new user similarity model to improve the accuracy of collaborative filtering. *Knowledge-Based Systems*, *56,* 156-166.

Lu, J., Shambour, Q., Xu, Y., Lin, Q., & Zhang, G. (2013). A WEB-BASED PERSONALIZED BUSINESS PARTNER RECOMMENDATION SYSTEM USING FUZZY SEMANTIC TECHNIQUES. *Computational Intelligence*, *29*(1), 37-69.

Massa, P., & Avesani, P. (2007). Trust-aware recommender systems. In Proceedings of the 2007 ACM conference on Recommender systems (pp. 17-24). ACM.

Moradi, P., & Ahmadian, S. (2015). A reliability-based recommendation method to improve trust-aware recommender systems. *Expert Systems with Applications*.

Polatidis, N., & Georgiadis, C. K. (2013). Recommender Systems: The Importance of Personalization in E-Business Environments. *International Journal of E-Entrepreneurship and Innovation (IJEEI)*, *4*(4), 32-46.

Polatidis, N., & Georgiadis, C. K. (2015). A ubiquitous recommender system based on collaborative filtering and social networking data. *International Journal of Intelligent Engineering Informatics,* 3(2/3), pp. 186-204.

Shani, G., & Gunawardana, A. (2011). Evaluating recommendation systems. In *Recommender systems handbook* (pp. 257-297). Springer US.

Shi, Y., Larson, M., & Hanjalic, A. (2014). Collaborative filtering beyond the user-item matrix: A survey of the state of the art and future challenges. *ACM Computing Surveys (CSUR)*, *47*(1), 3.

Son, L. H. (2014). HU-FCF: a hybrid user-based fuzzy collaborative filtering method in recommender systems. *Expert Systems with Applications*, *41*(15), 6861-6870.

Su, X., & Khoshgoftaar, T. M. (2009). A survey of collaborative filtering techniques. *Advances in artificial intelligence*, *2009*, 4.

Toledo, R. Y., Mota, Y. C., & Martínez, L. (2015). Correcting noisy ratings in collaborative recommender systems. *Knowledge-Based Systems*, *76*, 96-108.

Wang, W., Zhang, G., & Lu, J. (2015). Collaborative Filtering with Entropy-Driven User Similarity in Recommender Systems. *International Journal of Intelligent Systems 30(8), pp. 854-870*.